\documentclass[aps,prd,reprint,twocolumn,superscriptaddress,longbibliography,nofootinbib,floatfix,showpacs]{revtex4-1}
\usepackage{epsfig}
\usepackage{amsmath,amssymb,amsfonts}
\usepackage{hyperref}
\usepackage{mathrsfs}
\usepackage{bbm}
\usepackage{slashed}
\usepackage{graphicx}
\usepackage{verbatim}

\usepackage{bm}

\usepackage[usenames]{xcolor}

\definecolor{darkgreen}{rgb}{0.2,0.6,0}

\newcommand{\be}{\begin{equation}}
\newcommand{\ee}{\end{equation}}
\newcommand{\bw}{\begin{widetext}}
\newcommand{\ew}{\end{widetext}}
\newcommand{\bi}{\begin{itemize}}
\newcommand{\ei}{\end{itemize}}
\newcommand{\ud}{\mathrm{d}}

\newcommand{\LCm}{{\scriptscriptstyle -}} \newcommand{\LCp}{{\scriptscriptstyle +}}
\newcommand{\LCpm}{{\scriptscriptstyle \pm}}

\newcommand{\LCperp}{{\scriptscriptstyle \perp}}

\newcommand{\pa}{\partial}
\newcommand{\bo}{\textbf}
\renewcommand{\Im}{\, \text{Im} \,}

\usepackage[T1]{fontenc} \usepackage[latin1]{inputenc}

\begin{document}

\title{Momentum spectrum of Schwinger pair production in four-dimensional e-dipole fields}

\author{Gianluca Degli Esposti}
\email{g.degli-esposti@hzdr.de}
\affiliation{Helmholtz-Zentrum Dresden-Rossendorf, Bautzner Landstra{\ss}e 400, 01328 Dresden, Germany}

\affiliation{Institut f\"ur Theoretische Physik, 
Technische Universit\"at Dresden, 01062 Dresden, Germany}

\author{Greger Torgrimsson}
\email{greger.torgrimsson@umu.se}
\affiliation{Department of Physics, Ume{\aa} University, SE-901 87 Ume{\aa}, Sweden}

\begin{abstract}

We calculate the momentum spectrum of electron-positron pairs created via the Schwinger mechanism by a class of four-dimensional electromagnetic fields called e-dipole fields. To the best of our knowledge, this is the first time the momentum spectrum has been calculated for 4D, exact solutions to Maxwell's equations. Moreover, these solutions give fields that are optimally focused, and are hence particularly relevant for future experiments. To achieve this we have developed a worldline instanton formalism where we separate the process into a formation and an acceleration region. 

\end{abstract}
\maketitle

Schwinger pair production is challenging for both experiment and theory~\cite{Sauter:1931,Schwinger:1951,Dunne:2004nc,DiPiazza:2011tq,Gelis:2015kya,Fedotov:2022ely}. It requires field strengths much higher than what today's high-intensity-laser facilities can reach. And its nonperturbative nature makes it difficult to calculate the probability for physical, 4D fields. Collision of several pulses have been suggested as a way to reduce the required field strength~\cite{Bulanov:2010ei}. 
There is a class of fields called e-dipole fields~\cite{dipolePaper} which are exact solutions to Maxwell's equations and represent actual, physical fields that are optimally focused for Schwinger pair production~\cite{Gonoskov:2013ada}. They are genuinely 4D and hence computationally challenging. In principle, the probability (neglecting radiative corrections) is determined by solutions to the Dirac equation with a background field. But in practice, no one has managed to solve this numerically\footnote{See~\cite{Aleksandrov:2017mtq,Lv:2018wpn,Ababekri:2019dkl,Kohlfurst:2019mag} for state of the art.}. One therefore has to resort to approximations. We are interested in approximations for field strengths well below the Schwinger field\footnote{From now on we will use units with $c=\hbar=m=1$ and we absorb $e$ into the field strength of the background field, $eF_{\mu\nu}\to F_{\mu\nu}$. In particular, $E_S=1$.} $eE_S=m^2$. Indeed, the fields will likely be weak in the future experiments that detect this process for the first time. 

Much work has been done for special backgrounds such as fields which depend on only one spacetime coordinate \cite{Brezin:1970,Popov:1972,Popov:2005,Dunne:2005sx,Dunne:2006st}, using e.g. the Wentzel-Kramers-Brillouin (WKB) method. For spacetime fields, however, a generalization of the WKB method seems challenging, despite recent progress in 2D for colliding laser pulses \cite{Kohlfurst:2021skr}.

Apart from the maximum field strength, $E$, another relevant parameter is $\gamma=\omega/E$, where $\omega$ is some characteristic length scale, which can be defined in terms of the curvature of the field at the maximum. If $\gamma\ll1$ the probability integrated over all momenta and summed over spin can be approximated by (see e.g.~\cite{Bulanov:2004de,Dunne:2006st})
\be\label{standardLCF}
\mathbb{P}_{LCF}=2\int\ud^4x\frac{\mathcal{E}^2(x)}{(2\pi)^3}\exp\left(-\frac{\pi}{\mathcal{E}(x)}\right) \;,
\ee
where $\mathcal{E}=\sqrt{-F_{\mu\nu}F^{\mu\nu}/2}=\sqrt{{\bf E}^2-{\bf B}^2}$ (${\bf E}\cdot{\bf B}=0$ for e-dipole fields). This locally-constant-field (LCF) approximation was used in~\cite{Gonoskov:2013ada}. For $E\ll1$ one can perform the integrals in~\eqref{standardLCF} with the saddle-point method.

For $\gamma\sim1$ one cannot use~\eqref{standardLCF}. Instead, one can use a worldline-instanton formalism \cite{Affleck:1981bma,Dunne:2005sx,Dunne:2006st,Dunne:2006ur,Dumlu:2011cc,Gould:2017fve,Schneider:2018huk,Edwards:2019eby}.
In the usual approach, the integrated probability is obtained from the imaginary part of the effective action, which in turn is represented by a path integral over closed worldlines (i.e. loops, periodic in both space and time). It was shown in~\cite{Schneider:2018huk} how to use this formalism for 4D fields, in particular for an e-dipole field.  

However, neither~\eqref{standardLCF} nor the closed-worldline formalism give any information about the momentum or spin of the pair. In~\cite{DegliEsposti:2021its} we showed how to use open worldlines\footnote{Open worldlines have been used for pair production by a constant field in~\cite{Barut:1989mc,Rajeev:2021zae}.} to obtain the momentum spectrum for time-dependent fields, and in~\cite{DegliEsposti:2022yqw} we generalized to 2D fields, with a single electric component, no magnetic field, and which only depend on $t$ and $z$. Here we will for the first time calculate the spectrum of 4D fields, which are exact solutions to Maxwell's equations. We emphasized in~\cite{DegliEsposti:2021its,DegliEsposti:2022yqw} that the instantons are not unique because one is free to make a deformation of the complex proper-time contour without changing the probability. Here we show how to choose a contour which allows us to clearly separate the process into a formation region, where the instanton is complex and where the ``creation happens'', and a subsequent acceleration region, where the real particles are accelerated by the field. We are not trying to answer 
questions such as ``when are the particles actually created'', and we are not suggesting that one tries to place detectors inside the field\footnote{See~\cite{Ilderton:2021zej} for recent insight into the different definitions of time-dependent particle numbers.}. However, we will show that this contour gives an advantage both numerically and analytically.

A general e-dipole field is determined by~\cite{dipolePaper,Gonoskov:2013ada}
\be\label{edipoleFromZ}
{\bf Z}={\bf e}_z\frac{3E}{4r}[g(t+r)-g(t-r)] \;,
\ee
where $r=\sqrt{x^2+y^2+z^2}$ and $g$ is an arbitrary function. We focus here on symmetric fields with a single maximum. The fields are given by ${\bf E}=-\nabla\times\nabla\times{\bf Z}$ and ${\bf B}=-\nabla\times\partial_t{\bf Z}$. 
The probability amplitude is obtained with the Lehmann-Symanzik-Zimmermann (LSZ) reduction formula \cite{ItzyksonZuber,Barut:1989mc} ($px=p_\mu x^\mu$, $g_{\mu\nu}=\text{diag}(1,-1,-1,-1)$),
\be\label{LSZ3pair}
\begin{split}
M=\lim_{t_\LCpm\to\infty}\int \ud^3x_\LCp\ud^3 x_\LCm e^{ipx_\LCp+ip'x_\LCm}
\bar{u}\gamma^0S(x_\LCp,x_\LCm)\gamma^0 v \;,
\end{split}
\ee
where $u({\bf p})$ and $v({\bf p}')$ are free asymptotic electron and positron states, and $S$ is the background-field dependent fermion propagator, which can for an arbitrary background be expressed as a path integral over particle trajectories $q^\mu(\tau)$,
\be\label{propagatorWorldline}
\begin{split}
&S(x_\LCp,x_\LCm)=(i\slashed{\partial}_{x_\LCp}-\slashed{A}(x_\LCp)+1)\int_0^\infty\frac{\ud T}{2}\int\limits_{q(0)=x_\LCm}^{q(1)=x_\LCp}\mathcal{D}q\,\mathcal{P}\\
&\times\exp\left\{-i\left[\frac{T}{2}+\int_0^1\!\ud\tau\left(\frac{\dot{q}^2}{2T}+A\dot{q}+\frac{T}{4}\sigma^{\mu\nu}F_{\mu\nu}\right)\right]\right\} \;,
\end{split}
\ee
where $T$ is the total length of proper time, $\tau$ is proper time rescaled by $T$, $\mathcal{P}$ means proper-time ordering, and $\sigma^{\mu\nu}=\frac{i}{2}[\gamma^\mu,\gamma^\nu]$.
Since the field is 4D, all the integrals are nontrivial. We have performed them using the saddle-point method. The saddle point for the path integral is called a worldline instanton, and it is determined by the Lorentz-force equation, $\Ddot{q}^\mu = T F^{\mu \nu}\dot q_\nu$. For $T$ and ${\bf x}_\LCpm$ the saddle points are determined by $T^2 = \dot q^2$, $\dot q_i(1) = Tp_i$ and $\dot q_i(0) = -Tp'_i$, fixing the instanton in terms of the asymptotic momenta ${\bf p}$ and ${\bf p}'$, which are at this point free parameters. However, the peaks of the spectrum are simply Gaussian~\eqref{gaussianSpectrum}, which we can characterize uniquely by giving the widths and the integrated probability. To calculate these quantities we only need to find instantons, plus the solutions to the first-order variation of the Lorentz-force equation, for the saddle-point values of the momenta, ${\bf p}_s$ and ${\bf p}'_s$. Since $p_{s\LCperp} = p'_{s\LCperp} = 0$, where $p_\LCperp=\{p_x,p_y\}$ etc., the instanton follows the $z$ axis ($q^\LCperp(\tau) = 0$), on which $\bo{B} = 0$, $E_x = E_y = 0$, and the Lorentz-force equation reduces to a 2D problem, $\Ddot{t} = T E_3(t,z) \dot z$ and $\Ddot{z} = T E_3(t,z) \dot t$.
However, this does not mean that everything is the same as in the 2D case. Indeed, the spectrum in the 2D case does not even have the same number of independent momentum components, see e.g.~\eqref{gaussianSpectrum}.

\begin{figure}[!ht]
\centering
\includegraphics[width=\linewidth]{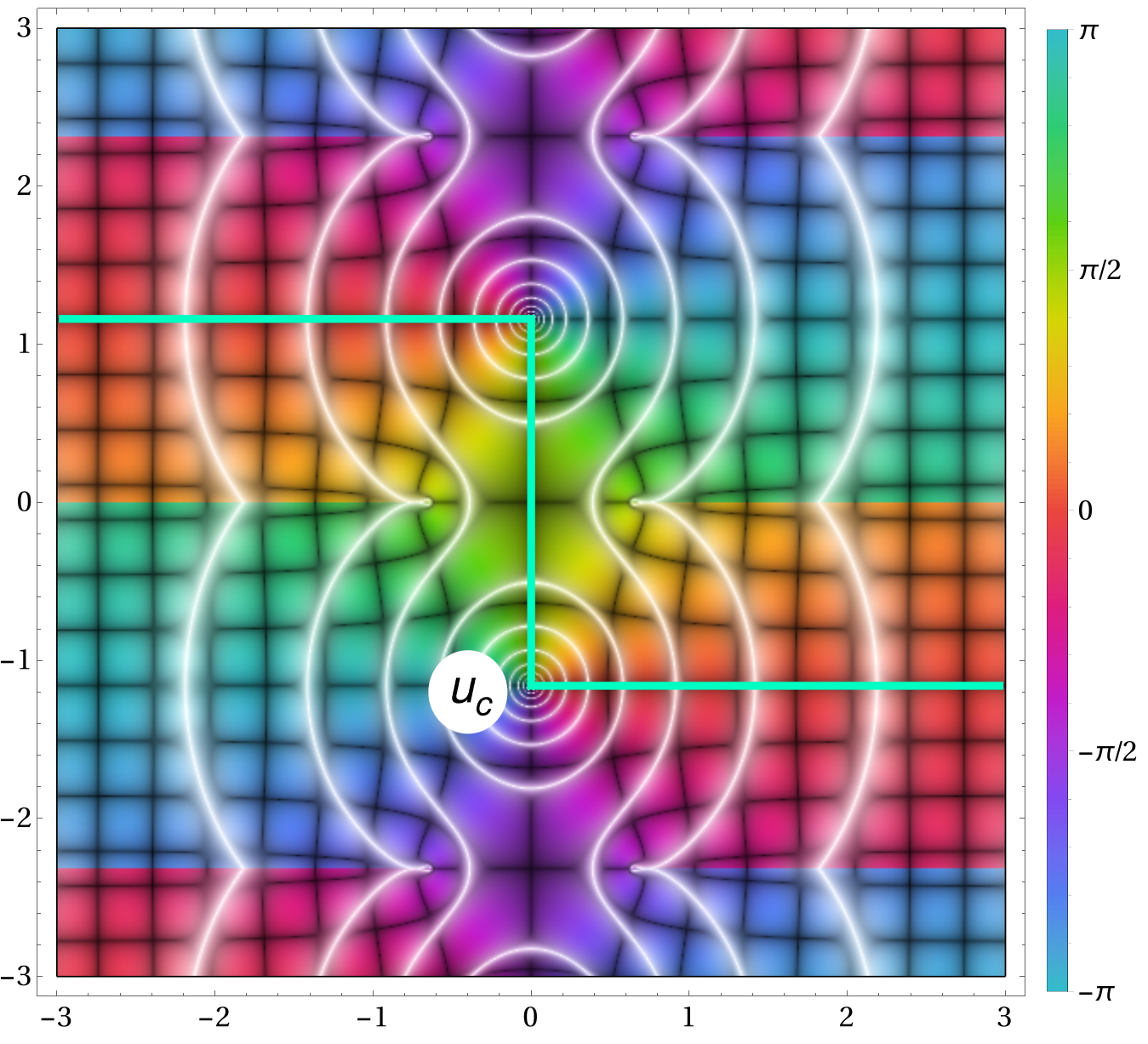}
\caption{$t(u)$ in the complex $u$ plane for $\gamma=1$. The color represents the phase, the white curves are contour lines of $|t(u)|$, and the black curves are lines of constant real/imaginary part. The green line shows our preferred contour. The details on how we obtained this plot are in Appendix~\ref{Complex plot}.}
\label{fig:instanton2}
\end{figure}

\begin{figure}[!ht]
\centering
\includegraphics[width=\linewidth]{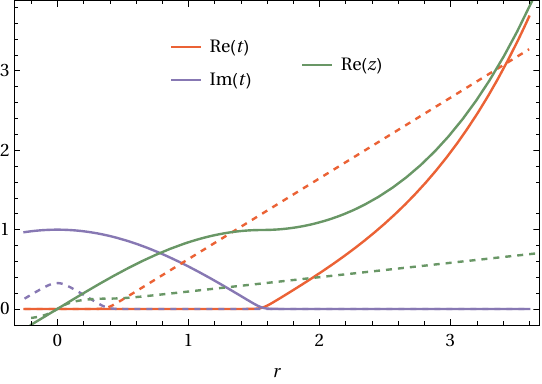}
\caption{Instantons for $\gamma = 1/10$ (solid line) and $\gamma = 5$ (dashed). We see that the size of the creation region is much smaller for large $\gamma$. At small $\gamma$ we see that the $t$ and $z$ components converge for large $r$.}
\label{fig:instanton}
\end{figure}

After having derived the saddle-point equations, it is more convenient to change variable from $\tau$ to $u = T(\tau - 1/2)$, so that the instanton obeys $q_\mu''={F_\mu}^\nu q_\nu'$, $q^{\prime2}=1$, $q_i'(u_1)=p_i$ and $q_i'(u_0) = -p'_i$, where $-u_0=u_1=T/2$. Since $T\to\infty$ as $t_\LCpm\to\infty$, $u$ starts at $-\infty$ and goes to $+\infty$. $T$ no longer appears in the EOM. We can think of $u=0$ as the start of the creation, and the half of the contour that goes to $+\infty$ ($-\infty$) describes the electron (positron). Since $t(u)$ is symmetric and $z(u)$ antisymmetric, the electron and the positron both propagate forward in time but in opposite directions along the $z$ axis.  
The contour for $u$ is complex, and we are free to make contour deformations. Although they give the same probability, they are not equally simple. We parametrize the contour as $u'(r)=f(r)$ where $r\in\mathbb{R}$. We have chosen $f(r)=1-(i+1)\psi(r)$, where $\psi\approx1$ for $|r|<r_c$ and $\psi\approx0$ for $|r|>r_c$, for some constant $r_c$. $u$ starts at $0$, follows the negative imaginary axis to $u_c=-i|u_c|$, turns and goes to $\infty$ parallel to the positive real axis, see Fig.~\ref{fig:instanton2}. Some parts of the instanton always have to be complex, regardless of the choice of contour. One might still expect the instanton to be real asymptotically, but this is not automatic, and is not the case for the contour we advocated in~\cite{DegliEsposti:2022yqw}. We can choose $r_c$ such that the instanton is real asymptotically, but $r_c$ will then depend on e.g. $\gamma$. Since we will find the same probability regardless of the contour, it might seem like unnecessary work trying to find such a $r_c$~\cite{DegliEsposti:2022yqw}. However, we will show that it is in fact useful for practical calculations.
As initial conditions at $u=0$ we have $z=t'=0$ from symmetry and $z'=i$ from $q^{\prime2}=1$. We then adjust the two constants $t(0)=i|t(0)|$ and $r_c$ until we find an instanton with $\text{Im }t(r_a)=\text{Im }t(r_b)=0$, for some arbitrary points $r_a,r_b>r_c$. 
The instanton will then be real for $r>r_c$ and describe the trajectory of real particles, see Fig.~\ref{fig:instanton}. Note, importantly, none of the conditions at $u=0$ or $r_{a,b}$ involves ${\bf p}$ or ${\bf p}'$. The solution will automatically be the instanton for the saddle-point values of ${\bf p}$ or ${\bf p}'$. After we have found the instanton we obtain the energy by simply evaluating $p_0=t'(\infty)$.  
We will call $|r|<r_c$ the formation region, where the creation happens, and $|r|>r_c$ the acceleration region. $t(u)$ and $z'(u)$ are imaginary (real) for $|r|<r_c$ ($|r|>r_c$), so $t(\pm u_c)=z'(\pm u_c)=0$, see Fig.~\ref{fig:instanton}. Thus, we can think of $u_c$ ($-u_c$) as the point where the electron (positron) goes from being a virtual to a real particle. The pair is created at $t=0$ with zero momentum. But $z(u_c)=-z(-u_c)\ne0$, so the electron and positron are created at different points in space.   
Thus, this choice of contour allows for a natural interpretation. 

More importantly, it is useful in practice. We cannot know what values of $\gamma$ will be be relevant in future experiments, but, judging from current laser facilities, one can guess $\gamma\ll1$. This is also the regime which is most Schwinger-like, since for $\gamma\gg1$ the production would instead be perturbative. For $\gamma\ll1$ we need to find the instantons up to very large $r$ to see convergence to the asymptotics, which means many numerical time steps. For example, for $\gamma=0.01$ we had to consider $r=\mathcal{O}(10^4)$. This is due to the fact that at $\gamma \ll 1$ the field is wide, and the electron (positron) travels at $z\approx t$ ($z\approx -t$) which affects the convergence of $g(t\pm z)$, so it takes longer for the particles to become free. But with the above choice of contour, $r_{a,b}$ do not need to be large, they just have to be larger than $r_c\approx\pi/2$. This is a huge advantage, because to find $t(0)$ and $r_c$ we solve the Lorentz-force equation many times, but only up to $r_{a,b}$, which is much faster than if we had used a different contour with conditions at $r\gg1$. After we have found $t(0)$ and $r_c$ we solve up to $r\gg1$, but we only have to do that once. We will show that this contour also helps in analytical calculations.       

To obtain the prefactor we expand the exponent to second order around the saddle points and perform the resulting Gaussian integrals, which give determinants of Hessian matrices. For the path integral this is done using the Gelfand-Yaglom method. See Appendix~\ref{Gelfand-Yaglom and the prefactor}.
We find
\be\label{eq:Probability}
\mathbb P= \int\frac{\ud^3p \, \ud^3p'}{(2\pi)^6}\mathbb{P}(p,p')
\quad
\mathbb{P}(p,p')=\frac{2(2\pi)^3e^{-\mathcal A}}{|h\bar{\phi}^{\prime2}|p_0 p'_0} \;,
\ee
where $\mathcal A = 2 \Im \int \ud u \, q^\mu \pa_\mu A_\nu \frac{\ud q^\nu}{\ud u}$, and $h$ and $\bar{\phi}$ are two functions coming from the Gelfand-Yaglom method.
Since the field is 4D, there are no volume factors and none of the components of the momentum is conserved.

\begin{figure*}[!ht]
\centering
\raisebox{-.1cm}{\includegraphics[width=.522\linewidth]{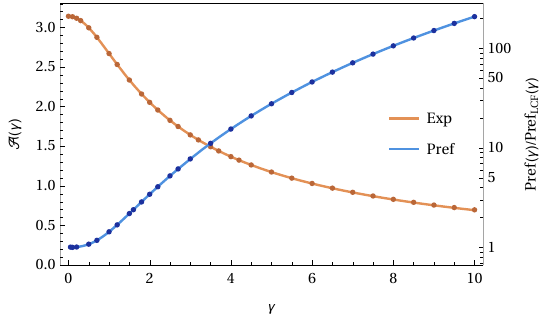}}
\hspace{0.2cm}
\includegraphics[width=.456\linewidth]{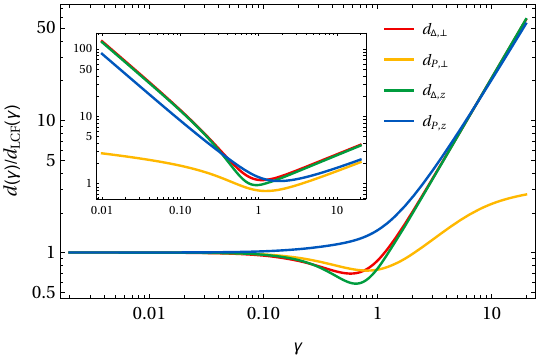}
\caption{Left: Comparison with the effective action method~\cite{Schneider:2018huk} (dots) for the exponent (without the overall factor of $1/E$) and the normalized prefactor. The number of points used for the discrete instantons is $N = 1000$. The same plots for the Lorentzian dipole can be found in appendix~\ref{LCF expansions in the formation region}. Right: Widths normalized by their LCF approximations and unnormalized (but without overall factor of $\sqrt{E}$). We see that the two $d_\Delta$ widths are very similar, with $d_{\Delta,\perp}$ being slightly bigger than $d_{\Delta,z}$.}
\label{fig:comparison}
\end{figure*}

To find the widths we change variables to $p_j = -P_j + \frac{\Delta p_j}{2}$ and $p'_j = P_j + \frac{\Delta p_j}{2}$.  
Due to symmetry there are only four nonvanishing independent widths and the spectrum has the form
\be\label{gaussianSpectrum}
\begin{split}
&\mathbb P(p,p')=\frac{2(2\pi)^3e^{-\mathcal{A}}}{|h\bar{\phi}^{\prime2}| p_0^2}\\
\times&\exp \left\{-\frac{\Delta p_\LCperp^2}{d_{\Delta,\perp}^2} -\frac{\Delta p_z^2}{d_{\Delta,z}^2} -\frac{P_\LCperp^2}{d_{P,\perp}^2} -\frac{(P_z -\mathscr P)^2}{d_{P,z}^2} \right\} \;,
\end{split}
\ee
where from now on $\mathcal{A}=\mathcal{A}({\bf p}_s,{\bf p}'_s)$ and $P_\LCperp^2=P_x^2+P_y^2$ etc.
To obtain the widths we need to solve
\be\label{Jacobi}
(- g_{\mu\nu}\pa_u^2 + F_{\mu \nu} \pa_u + q'^\rho \pa_\nu F_{\mu \rho})\delta q^\nu(u)=0 \;,
\ee
which comes from expanding the Lorentz-force equation around the instanton for ${\bf p}_s$, ${\bf p}'_s$. The equation for $\delta x$ and $\delta y$ are the same. $\delta t$ and $\delta z$ are combined into a single variable, $\eta$. We find (see Appendix~\ref{Local-nonlocal separation}) 
\be\label{deltaxeq}
\begin{split}
    \eta'' &= (E^2+\nabla E\cdot\{z',t'\})\eta \\
    \delta x'' &= (t' \pa_x E_x - z' \pa_x B_y) \, \delta x =-\frac{1}{2}\nabla E\cdot\{z',t'\}\delta x \;,
\end{split}
\ee
where $\nabla E=(\partial_t E_3,\partial_z E_3)$. 
Note that the magnetic field contributes to $\delta x$, but can be replaced since Maxwell's equations plus symmetry imply $\pa_x E_x = \pa_y E_y = -\frac{1}{2} \pa_z E_z$, $\pa_x B_y = -\pa_y B_x =\frac{1}{2}\pa_t E_z$. 
The initial conditions are
\be\label{initialConditionsetadelta}
\begin{split}
\eta_a(0)&=0 \quad \eta_a'(0)=1 
\quad
\eta_s(0)=1 \quad \eta_s'(0)=0 \\
\delta x_a(0)&=0 \quad \delta x_a'(0)=1 
\quad
\delta x_s(0)=1 \quad \delta x_s'(0)=0 \;.
\end{split}
\ee
For a general contour we have $d_{\Delta,z}^{-2}=\frac{1}{2p_0^2} \Im \left( \frac{t}{p_0}-\frac{\eta_a}{\eta_a'}\right)$, and similar for the other widths, see Appendix~\ref{Local-nonlocal separation}. With our choice of contour we can rewrite these as
\be\label{dFromW}
\begin{split}
d_{\Delta,\perp}^{-2}&=\frac{W(\delta x_{sr},\delta x_{si})}{2|\delta x_s'|^2}
\qquad
d_{P,\perp}^{-2}=2\frac{W(\delta x_{ar},\delta x_{ai})}{|\delta x_a'|^2} \\
d_{\Delta,z}^{-2}&=\frac{W(\eta_{ar},\eta_{ai})}{2p_0^2|\eta_a'|^2} 
\qquad
d_{P,z}^{-2}=2\frac{W(\eta_{sr},\eta_{si})}{p_0^2|\eta_s'|^2} \;,
\end{split}
\ee
where $W(f,g)=fg'-f'g$ is the Wronskian, $\eta_{ar}=\text{Re }\eta_a$ and $\eta_{ai}=\text{Im }\eta_a$ etc., and where all quantities are evaluated at $u\to\infty$. Outside the formation region, $\eta_{ar}$ and $\eta_{sr}$ are separately solutions to~\eqref{deltaxeq}, so $(\ud/\ud r) W(\eta_{ar},\eta_{ai})=0$ for $r>r_c$.
Hence, the Wronskians can be evaluated at $u\gtrsim u_c$, rather than at $u\to\infty$, and are therefore local contributions to the widths. $|\eta'|^2$ and $|\delta x'|^2$ are not constant for $r>r_c$ and are therefore nonlocal contributions. We also have $|h(\infty)|=2|\eta_s'\eta_a'|$ and $|\bar{\phi}'(\infty)|=2|\delta x_s'\delta x_a'|$, see Appendix~\ref{Local-nonlocal separation}.
We find
\be\label{PrefWronskians}
\mathbb{P}=\frac{[W(\eta_{ar},\eta_{ai})W(\eta_{sr},\eta_{si})]^{-1/2}e^{-\mathcal{A}}}{32W(\delta x_{ar},\delta x_{ai})W(\delta x_{sr},\delta x_{si})} \;.
\ee
All nonlocal contributions have canceled. Thus, the integrated probability only depends on the part of the field that $q^\mu$ and $\delta q^\mu$ ``see'' while $|r|<r_c$. This provides further motivation for calling $|r|<r_c$ the formation region, because it agrees with the intuition that the integrated probability should not depend on what happens with the particles after they have been created.   

We allow $\gamma=\mathcal{O}(1)$, so in general the instantons etc. have a complicated dependence on $\gamma$. But $E\ll1$ is the expansion parameter, and nothing will have any nontrivial dependence on $E$. To make this clear right from the start, we rescale $q^\mu\to q^\mu/E$ and $u\to u/E$, so $E$ no longer appears in the Lorentz-force equation or any other EOM. We have $\mathcal{A}\propto1/E$ and, for all widths, $d\propto\sqrt{E}$.

We can compare the integrated probability~\eqref{eq:Probability} with the closed-instanton method in~\cite{Schneider:2018huk}. Fig.~\ref{fig:comparison} shows the results for a Gaussian pulse, $g'''(t) = e^{-\omega^2 t^2}$.  
We find perfect agreement.

The local-nonlocal separations is also useful for deriving $\gamma\ll1$ approximations. The Wronskians only depend on the formation region, where we can expand the instanton, $\eta$ and $\delta x$ as sums of $\mathcal{O}(1)$ and $\mathcal{O}(\gamma^2)$ terms. 
These expansions of $q$, $\eta$ and $\delta x$ are given in Appendix~\ref{LCF expansions in the formation region}.
We find $W(\eta_{ar},\eta_{ai})\approx\frac{\pi}{10}\gamma^2$, $W(\eta_{sr},\eta_{si})\approx\frac{\pi}{2}\gamma^2$, $W(\delta x_{sr},\delta x_{si})\approx\frac{\pi}{5}\gamma^2$ and $W(\delta x_{ar},\delta x_{ai})\approx\frac{\pi}{2}$.
Inserting this into~\eqref{PrefWronskians} gives $\mathbb{P}\approx\frac{5\sqrt{5}}{(2\pi)^3\gamma^4}e^{-\pi/E}$, which agrees with what one finds by performing the integrals in~\eqref{standardLCF} with the saddle-point method. 

The nonlocal parts, $|\eta'|$ and $|\delta x'|$, are more challenging. Here we cannot expand $t$ and $z$ as a power series in $\gamma$, since $\gamma t,\gamma z=\mathcal{O}(1)$ in the acceleration region, as expected since the momentum spectrum depends on how the field accelerates the particles after they have been created and until they leave the field.  
We first note that $\gamma\ll1$ means a very wide field, so compared to the length scale of the field, the particles are quickly accelerated to highly relativistic velocities. The instanton will therefore follow almost lightlike trajectories, $z\approx t$, see Fig.~\ref{fig:instanton}. It is therefore convenient to use lightfront coordinates, $\phi=\frac{\gamma}{2}(t+z)$ and $\theta=\gamma(t-z)$.
One of the two nonzero Lorentz-force equations becomes $\phi''=F(\phi,\theta)\phi'$.
The other, $\theta''=-F\theta'$, can be replaced by the on-shell condition $(t')^2-(z')^2=1$, which gives $\theta'=\frac{\gamma^2}{2\phi'}$, with $\theta(0)=i\gamma$.
In the formation region we have $F\approx1$, while in the acceleration region $F(\phi,\theta)\approx F(\phi,0)=:F(\phi)$. In both regions we therefore have $\phi\approx\phi_0$ where $\phi''_0=F(\phi_0)\phi_0'$.
There are no explicit factors of $\gamma$ in this equation, but there are in the initial conditions $\phi_0(0)=\phi_0'(0)=i\gamma/2$, and $\phi_0'(u)\approx H(\phi_0)$, where $H(x)=\int_0^x\!\ud\varphi\, F(\varphi)$.
Thus, the asymptotic momentum is $p_0=t'(\infty)\approx H(\infty)/\gamma=\mathcal{O}(1/\gamma)$.

The derivations of $\eta_{a,s}'(\infty)$ and $\delta x_{a,s}'(\infty)$ are quite long, see Appendix~\ref{The longitudinal widths} and~\ref{The transverse widths}. The results for $\eta$, however, are very simple, $p_0^2|\eta_a'(\infty)|^2\approx\frac{1}{4}$, $p_0^2|\eta_s'(\infty)|^2\approx\frac{9}{4}$ and $p_0^2|h|\approx\frac{3}{2}$.
$\delta x_{a,s}$ are nontrivial. $\delta x_s$ is first obtained by changing variables from $u$ to $\phi$ and solving $H\delta x''(\phi)+F\delta x'(\phi)=-\frac{1}{2}F'(\phi)\delta x$
with initial conditions $\delta x_s(\phi=0)=1$ and $\delta x_s'(\phi=0)=0$. Thus, $\delta x_s$ is independent of $\gamma$ to leading order. This gives $\delta x_s'(u=\infty)=H(\infty)\delta x_s'(\phi=\infty)$. $\delta x_a$ is obtained from $\delta x_s$ using Abel's identity, which gives
\be
\begin{split}
\frac{\delta x_a'(\infty)}{\delta x_s'(\infty)}&\approx c_1\ln\left(\frac{1}{\gamma}\right)+c_2
=-\ln\left(a\frac{\gamma}{2}\right)-\frac{i\pi}{2}\\
&+\int_0^\infty\ud\phi\left(\frac{1}{H\delta x_s^2}-\frac{1}{\phi(1+a\phi)}\right) \;,
\end{split}
\ee
where $a$ is an arbitrary constant. Convergence to this LCF approximation of the widths is demonstrated in Fig.~\ref{fig:comparison}. 
Thus $d_{\Delta,z},d_{P,z}, d_{\Delta,\perp}\propto\sqrt{E}/\gamma$, while $d_{P,\perp}\propto\sqrt{E}|c_1\ln(1/\gamma)+c_2|$.

The scaling of $d_{\Delta,\perp}$ suggests that it might be possible to produce particles with large $p_\LCperp, p_\LCperp'$, which could help to enhance $\chi=\sqrt{-(F^{\mu\nu}p_\nu)^2}$, which is otherwise small since $\chi=E\sqrt{1+p_\LCperp^2}$ for $x=y=0$. 
For $\chi\sim1$ the pair could emit hard photons, which could lead to further particle production, or even cascades~\cite{Bell:2008zzb,Fedotov:2010ja,Bulanov:2010gb,Nerush:2010fe,GrismayerCascades,Seipt:2020uxv,Fedotov:2022ely}.     
Even if no hard photons are emitted, one might still wonder if radiation reaction (RR) could be important for the spectrum. We show in Appendix~\ref{RR} that RR is negligible for ${\bf p}_s$ and ${\bf p}'_s$. 

We emphasize that for a 2D field, $E_3(t,z)$, one would have $d_{\Delta,\perp}=0$ due to momentum conservation. So the spectrum for a 2D field gives nothing with which one could even try to approximate $d_{\Delta,\perp}$. Moreover, we see in Fig.~\ref{fig:comparison} that $d_{\Delta,\perp}$ is not small, it is on the same order of magnitude as $d_{\Delta,z}$ and $d_{P,z}$. For a 1D field, $E_3(t)$, one would also have $d_{\Delta,z}=0$, but Fig.~\ref{fig:comparison} also shows that $d_{\Delta,z}$ too is not small.     

To conclude, we have for the first time calculated the momentum spectrum of pairs produced via the Schwinger mechanism by 4D solutions to Maxwell's equations. To do so we have developed a worldline instanton approach, which allows us to separate the process into a formation region, where the creation happens, and a subsequent acceleration region, where the real particles are accelerated to their final momentum. This is not only an intuitive picture, but is also useful in practice for both numerical and analytical calculations.    
These methods also pave the way for further investigations of other 4D fields, e.g. ones with more than one maximum, which leads to interference effects in the spectrum, and of nonlinear Breit-Wheeler pair production in 4D fields.

\acknowledgements

We are grateful to Christian Schneider for giving us a copy of his closed-worldline-instanton code, which we used to compare our results in Fig.~\ref{fig:comparison} 
G. T. is supported by the Swedish Research Council, Contract No. 2020-04327.

\appendix

\section{e-dipole fields}\label{Dipole fields}

The fields of an e-dipole can be obtained from ${\bf Z}$ in~\eqref{edipoleFromZ}, but this is not a gauge potential. As a gauge potential we can choose ${\bf A}=-\partial_t{\bf Z}$ (where $\{0,0,1\}\cdot{\bf A}=-A_3$ etc.), and with a corresponding nonzero $A_0$. For ${\bf Z}=Z(t,r){\bf e}_3$, we can write the gauge as
\be
A_\mu=\{\partial_z Z,0,0,\partial_t Z\} \;.
\ee
This automatically satisfies the Lorentz gauge condition $\partial_\mu A^\mu=0$.

Two pulse functions that differ by a second-order polynomial, 
\be\label{g1minusg2}
g_1(t)-g_2(t)=a+bt+ct^2 \;,
\ee
give the same electromagnetic field. We can therefore without loss of generality choose e.g.
\be\label{simplerg}
g(0)=g'(0)=g''(0)=0 \;,
\ee
or choose $g(t)$ such that it has no terms that go like $a+bt+ct^2$ for $t\to\infty$.

On the axis $x=y=0$ we have
\be
\begin{split}
E_3(t,z)=&\frac{3E}{2z^3}\{g(t-z)-g(t+z)\\
&+z[g'(t-z)+g'(t+z)]\} \;.
\end{split}
\ee
and $E_3(t,z=0)=Eg'''(t)$.
After rescaling $q^\mu\to q^\mu/E$ and $u\to u/E$, nothing depends nontrivially on $E$. We will use $F(t,z)=E_3(t,z)/E$ and $g(u)=G(\omega u)/\omega^3$, so 
\be\label{Fdef}
\begin{split}
F=&\frac{3}{2(\gamma z)^3}\{G[\gamma(t-z)]-G[\gamma(t+z)]\\
&+\gamma z(G'[\gamma(t-z)]+G'[\gamma(t+z)])\} \;.
\end{split}
\ee

In the $\gamma\ll1$ limit it is convenient to use lightfront coordinates,
\be\label{phitheta}
\phi=\frac{\gamma}{2}(t+z)
\qquad
\theta=\gamma(t-z) \;,
\ee
and $F(\phi)=F(\phi,\theta=0)$ is important for the leading order. For an e-dipole field we have
\be
F(\phi)=\frac{3}{2\phi^3}(-G[2\phi]+\phi G'[2\phi])=\frac{\ud}{\ud\phi}\frac{3G(2\phi)}{(2\phi)^2} \;,
\ee
where we have chosen $G$ as in~\eqref{simplerg}. This can be inverted 
\be\label{GfromH}
G(x)=\frac{x^2}{3}H\left(\frac{x}{2}\right) \;,
\ee
where
\be\label{HfromF}
H(x)=\int_0^x\!\ud\varphi\, F(\varphi) \;.
\ee
As mentioned in the main text, $H$ gives to leading order in $\gamma\ll1$ the energy as a function of lightfront time, $t'\approx\phi'/\gamma\approx H(\phi)/\gamma$. 
The field for Fig.~\ref{fig:comparison} was chosen to have a simple $E_z(t,x=y=z=0)$, but to simplify the calculation for $\gamma\ll1$ one could instead choose a simple $F(\phi)$, and then~\eqref{HfromF} and~\eqref{GfromH} give the corresponding $G$ (or $g$).
We can perform the integral in~\eqref{HfromF} using partial integration, which gives
\be\label{Hinfty}
H(\infty)=\int_0^\infty\!\ud\phi\,F(\phi)=3\int_0^\infty\!\ud\phi\,G^{(3)}(2\phi)=\frac{3}{2}G''(\infty) \;.
\ee
For example, for the Gaussian pulse $g'''(t) = e^{-\omega^2 t^2}$ we have $3G''(\infty)/2=3\sqrt{\pi}/4$.

\section{Gelfand-Yaglom and the prefactor}\label{Gelfand-Yaglom and the prefactor}

Evaluating the exponent at the saddle points one finds exactly the same result as in the time-dependent and 2D case.
As to the prefactor, we begin with the path integral using the Gelfand-Yaglom method. Expanding the exponent up to second order in $\delta q=q-q_{\text{inst}}$ gives 
\be
\exp\left\{-\frac{i}{2T}\int_0^1\delta q\Lambda\delta q\right\} \;,
\ee
where
\be\label{LambdaGeneral}
\Lambda_{\mu \nu} = T^2(- \eta_{\mu \nu} \, \pa_u^2 + F_{\mu \nu} \pa_u + q'^\rho \pa_\nu F_{\mu \rho}) \;,
\ee
which can be written in a block-diagonal form
\be
\Lambda =
\begin{pmatrix}
    \Lambda_{2D} & 0 & 0 \\
    0 & \Lambda_\perp & 0 \\
    0 & 0 & \Lambda_\perp
\end{pmatrix} \;,
\ee
where $\Lambda_{2D}$ is the $(t,z)$ block identical to the $2D$ case and
\be
\Lambda_\perp = T^2(\pa^2_u -t' \pa_x E_x + z'\pa_x B_y) \; .
\ee
This is a great simplification because the determinant splits 
\be
\det \Lambda = \det \Lambda_{2D} \, (\det \Lambda_\perp)^2
\ee
into the known $(t,z)$ contribution and a simpler factor 
\be
\det \Lambda_\perp=\phi(u_1) \;,
\ee
where $\phi$ is obtained by solving 
\be
\Lambda_\perp \phi = 0
\ee
with initial conditions 
\be\label{initialphi}
\phi(u_0) = 0
\qquad
\phi'(u_0) = 1/T \;,
\ee
see e.g.~\cite{Dunne:2007rt}. In order to take the asymptotic limit and show that factors of $t_\LCpm,T\to\infty$ cancel, we follow the treatment of $\Lambda_{2D}$ in~\cite{DegliEsposti:2022yqw}. We define $(\tilde u_0, \tilde u_1)$ such that it contains the interval where the field is not negligible and where the dynamics is nontrivial. $\tilde{u}_0$ and $\tilde{u}_1$ do not depend on $t_\LCpm$. We separate out the simple contribution coming from ``before'' $\tilde{u}_0$ (since the contour in $u$ is complex, we cannot simply express this as $u<\tilde{u}_0$) by noting that 
\be
\phi(\tilde u_0) \sim \frac{\tilde u_0 - u_0}{T} \sim \frac{t_\LCm}{Tp'_0}
\ee
and by defining $\phi = t_\LCm \, \bar \phi/(Tp'_0)$ so that $\bar \phi$ has initial conditions 
\be\label{initialbarphi}
\bar \phi(\tilde u_0) = 0
\qquad
\bar \phi'(\tilde u_0) = 1 \;,
\ee
which are independent of $t_\LCpm$. We can similarly separate out the contribution from after $\tilde{u}_1$ using $\phi(u_1)\approx \phi'(\tilde{u}_1)(u_1-\tilde{u}_1)$. Thus,
\be
\det \Lambda_\perp=\phi(u_1)\approx\bar \phi'(\tilde u_1)(u_1 -\tilde u_1)\frac{t_\LCm}{Tp'_0}\approx \bar \phi'(\tilde u_1) \frac{t_\LCm t_\LCp}{Tp'_0 p_0} \; .
\ee
$\bar \phi'(\tilde u_1)$ does not depend on $t_\LCpm$.
We can replace ``$\approx$'' with ``$=$'' in the asymptotic limit $t_\LCpm\to\infty$ and provided $\tilde{u}_0$ and $\tilde{u}_1$ are chosen large enough for a given precision goal (we consider in general fields such as $e^{-x^2}$ which are strictly speaking nonzero even asymptotically).

We perform the integrals over the ordinary variables as in~\cite{DegliEsposti:2022yqw}. 
Denoting the exponential part of the integrand as $e^{\varphi}$, we have
\be\label{derX1}
\begin{split}
\frac{\partial\varphi}{\partial x_\LCm^j}&=i[p'_j-q^{j\prime}(u_0)] 
\qquad
\frac{\partial\varphi}{\partial x_\LCp^j}=i[p_j+q^{j\prime}(u_1)] \\
\frac{\partial\varphi}{\partial T}&=\frac{i}{2}(a^2-1) \;,
\end{split}
\ee
where $a^2=q^{\prime2}$. In the limit $t_\LCpm\to\infty$ we have
\be
\begin{split}
q^{j\prime}(u_0)&=-\frac{x_\LCm^j}{T}\left(1+\frac{\sqrt{x_\LCp^2}}{\sqrt{x_\LCm^2}}\right) \\
q^{j\prime}(u_1)&=\frac{x_\LCp^j}{T}\left(1+\frac{\sqrt{x_\LCm^2}}{\sqrt{x_\LCp^2}}\right) \\
a^2&=\frac{\sqrt{x_\LCm^2}+\sqrt{x_\LCp^2}}{T} \;,
\end{split}
\ee
where $x_\LCpm^2=t_\LCpm^2-{\bf x}_\LCpm^2$. Denoting $\bo X=\{T, \bf x_\LCm,\bf x_\LCp \}$, the above equations give us $\partial\varphi/\partial X_j$, $j=1,\dots,7$, expressed explicitly in terms of ${\bf X}$. Solving $\partial\varphi/\partial X_j=0$ gives us the saddle point ${\bf X}_s$,
\be
x_{\LCm s}^j=-\frac{p'_j}{p'_0}t_\LCm
\quad
x_{\LCp s}^j=-\frac{p_j}{p_0}t_\LCp
\quad
T_s=\frac{t_\LCp}{p_0}+\frac{t_\LCm}{p'_0} \;.
\ee
Expanding the exponent to second order in ${\bf \delta X}={\bf X}-{\bf X}_s$ gives 
\be
\int \ud^7 \bo X\, \exp\{- \delta \bo X \cdot \bo H \cdot \delta \bo X\}=\sqrt{\frac{\pi^7}{\det \bo H}} \;,
\ee
where
\be
H_{ij}=-\frac{1}{2}\frac{\partial^2\varphi}{\partial X_i\partial X_j} \;.
\ee
Using Mathematica, it is straightforward to calculate ${\bf H}$, 
evaluate it at ${\bf X}_s$ and calculate the determinant. $\bo H$ itself does not have a simple form, but the determinant is (up to a phase)
\be
\det \bo H =\frac{p_0^5 p_0^{\prime 5}}{2^7 t_\LCm^3 t_\LCp^3 T} \;.
\ee 

Since we can evaluate the prefactor at the saddle point for the momenta, the $x$ and $y$ components of the instanton are zero, so $E_x=E_y=0$ and ${\bf B}=0$. This means the spin part is exactly the same as in the 2D case, so we can reuse the result in Eq.~(85) in~\cite{DegliEsposti:2022yqw}.   
Thus, the magnetic component does not have any effect on the spin structure for these fields.

Combining these contributions we find
\be\label{Pcombined}
\begin{split}
\mathbb P &= \int\frac{\ud^3p}{(2\pi)^3}\frac{\ud^3p'}{(2\pi)^3} 2p_0 p'_0 \bigg| \frac{e^{\dots}}{(2\pi T)^2} \sqrt{\frac{1}{\det \Lambda}} \sqrt{\frac{\pi^7}{\det \bo H}} \bigg|^2 \\
&= \int\frac{\ud^3p \, \ud^3p'}{(2\pi)^3} \frac{2}{|h(\tilde u_1)| \; |\bar \phi'(\tilde u_1)|^2 \; p_0 p'_0} \; e^{-\mathcal A}
\end{split}
\ee
with
\be\label{Adef}
\mathcal A = 2 \Im \int \ud u \, q^\mu \pa_\mu A_\nu \frac{\ud q^\nu}{\ud u} \; .
\ee
Since we can evaluate the prefactor at the momentum saddle point, we could replace $p_0'=p_0$ in the denominator in~\eqref{Pcombined}.

\section{Derivation of the widths}\label{Local-nonlocal separation}

In terms of 
\be
p_j = -P_j + \frac{\Delta p_j}{2}
\qquad
p'_j = P_j + \frac{\Delta p_j}{2}
\ee
we have a saddle point for the momentum variables at $\Delta p_j=0$ and $P_j=\delta_{j3}\mathscr{P}$. We start with the $\Delta p_j$ integrals. Expanding the exponent around the saddle point gives
\be\label{HessianForDeltap}
e^{-\mathcal{A}(\Delta p)}\to\exp\left\{-\mathcal{A}(0)-\frac{1}{2}\Delta p_i\frac{\partial^2\mathcal{A}}{\partial\Delta p_i\partial\Delta p_j}\Delta p_j\right\} \;.
\ee
We first calculate $\partial\mathcal{A}/\partial p_i$ and $\partial\mathcal{A}/\partial p'_i$ by going back to the exponent expressed as in~\eqref{LSZ3pair} and~\eqref{propagatorWorldline}, but now with $q^\mu$, $T$, and $\bf x_\LCpm$ replaced by their saddle-point values. These saddle points depend on ${\bf p}$ and ${\bf p}'$, but it follows from the definition of the saddle points that all first derivatives with respect to $q^\mu$, $T$, $\bf x_\LCpm$ vanish. The total derivatives with respect to ${\bf p}$ and ${\bf p}'$ are therefore equal to the partial derivatives, so we find
\be
\frac{\partial\mathcal{A}}{\partial p_j}=2\lim_{u\to\infty}\text{Im}\left(q^j+\frac{p_j}{p_0}t\right)
\ee
and
\be
\frac{\partial\mathcal{A}}{\partial p'_j}=2\lim_{u\to-\infty}\text{Im}\left(q^j+\frac{p'_j}{p'_0}t\right) \;.
\ee
Hence,
\be\label{dAdDeltap}
\frac{\partial\mathcal{A}}{\partial \Delta p_j}=\lim_{u\to\infty}\text{Im}\left(q^j+\frac{p_j}{p_0}t\right)+\lim_{u\to-\infty}\text{Im}\left(q^j+\frac{p'_j}{p'_0}t\right) \;.
\ee
For~\eqref{HessianForDeltap} we need the first derivative of~\eqref{dAdDeltap}, so when we expand the instanton around $\Delta p_j=0$ we only need the first-order variation, 
\be
q^\mu \to q^\mu + \Delta p_{j} \, \delta q^\mu_{(j)}+\mathcal{O}(\Delta p^2) \;,
\ee
which is determined by
\be\label{deltaqeq}
\frac{d^2}{du^2} \delta q^\mu_{(j)} = F^{\mu \nu} \frac{d}{du} \delta q_{(j),\nu} + \pa_\rho F^{\mu \nu} \, q'_\nu \, \delta q^\rho_{(j)} \;.
\ee
Note that this can be written as $\Lambda q=0$, where $\Lambda$ is the Hessian matrix for the worldline path integral~\eqref{LambdaGeneral}.
The boundary conditions $q_j'(-\infty)=-p_j'$ and $q_j'(+\infty)=p_j$ imply
\be\label{initialDeltaij}
\delta q'^i_{(j)}(\pm\infty)=\mp\frac{\delta_{ij}}{2} 
\qquad
\delta t'_{(j)}(\pm\infty) = -\frac{P_j}{2p_0} \;.
\ee
Because of symmetry, the term at $u=-\infty$ is equal to the one at $u=+\infty$, and we find
\be\label{ADeltaij}
\begin{split}
&\mathcal A^\Delta_{ij} := \frac{1}{2}\frac{\pa^2 \mathcal A}{\pa \Delta p_i \, \pa \Delta p_{j}} \\
&= \Im \left[ \delta q^i_{(j)} - \delta t_{(j)} \frac{P_i}{p_0} + \frac{t}{2p_0}\left(\delta^i_j- \frac{P_i P_j}{p_0^2} \right) \right](\infty) \;.
\end{split}
\ee

Since the $x$ and $y$ components of the instanton vanish, we only need the field and its derivatives evaluated at $x=y=0$, where $E_x=E_y=0$ and ${\bf B}=0$. The nonzero derivatives are
\be
\begin{split}
\pa_x E_x &= \pa_y E_y = -\frac{1}{2} \pa_z E_z\\
\pa_x B_y &= -\pa_y B_x =\frac{1}{2}\pa_t E_z \;.
\end{split}
\ee
The equations for $\delta x$ and $\delta y$ are the same,
\be\label{deltaxj}
\delta x'' = (t' \pa_x E_x - z' \pa_x B_y)\delta x =-\frac{1}{2}\nabla E\cdot\{z',t'\}\delta x \;,
\ee
where $\nabla E=\{\partial_t E_3,\partial_z E_3\}$.
An arbitrary solution to~\eqref{deltaxj} can be expressed as a superposition
\be
\delta x(u)=c_a\delta x_a(u)+c_s\delta x_s(u) \;,
\ee
where $\delta x_a$ and $\delta x_s$ are antisymmetric and symmetric solutions with initial conditions
\be
\delta x_a(0)=0 \quad \delta x_a'(0)=1 
\quad
\delta x_s(0)=1 \quad \delta x_s'(0)=0 \;.
\ee
For $j\ne1$ we have from~\eqref{initialDeltaij} $\delta x_{(j)}'(\pm\infty)=0$, but
since $\delta x_{a,s}'(\infty)\ne0$, this implies $\delta x_{(j)}(0)=0$. Thus, only $\delta x_{(1)}$ (and $\delta y_{(2)}$) is nonzero and is given by
\be
\delta x_{(1)}(u)=-\frac{1}{2}\frac{\delta x_s(u)}{\delta x_s'(\infty)} \;.
\ee
Substituting into~\eqref{ADeltaij} gives
\be
d_{\Delta,\perp}^{-2}=\mathcal A^\Delta_{11}=\mathcal A^\Delta_{22}=\frac{1}{2} \Im \left(\frac{t}{p_0} -\frac{\delta x_s}{\delta x_s'}\right)(\infty) \;.
\ee

For $\delta t_{(j)}$ and $\delta z_{(j)}$ we have initially two coupled equations,
\be\label{dtdzeq}
\begin{split}
\delta t''&=E\delta z'+\nabla E\cdot\{\delta t,\delta z\}z'\\
\delta z''&=E\delta t'+\nabla E\cdot\{\delta t,\delta z\}t' \;.
\end{split}
\ee
We can simplify this into a single relevant equation by replacing $\delta t$ and $\delta z$ with two new variables, $\eta$ and $\chi$, as in~\cite{DegliEsposti:2022yqw},
\be
\{\delta t,\delta z\}=\{t',z'\}\chi
+\{-z',t'\}\frac{\eta}{t^{\prime2}+z^{\prime2}} \;,
\ee
where $\eta=t'\delta z-z'\delta t$ is the relevant parameter.
Instead of~\eqref{dtdzeq} we have
\be\label{etaEqij}
\eta''=(E^2+\nabla E\cdot\{z',t'\})\eta
\ee
and
\be
\chi'=E\frac{1-4t^{\prime2}z^{\prime2}}{(t^{\prime2}+z^{\prime2})^2}\eta+\frac{2t'z'}{t^{\prime2}+z^{\prime2}}\eta' \;.
\ee
Note that the equation~\eqref{etaEqij} for $\eta$ does not involve $\chi$.
With the asymptotic condition for the instanton, $t'(\infty)=p_0$ and $z'(\infty)=P$, we can rewrite the contribution to~\eqref{ADeltaij} as
\be
\Im \left[\delta z_{(j)} - \delta t_{(j)} \frac{P}{p_0}\right](\infty)=\Im\left[\frac{\eta_{(j)}(\infty)}{p_0}\right] \;.
\ee
Thus, $\chi$ does not contribute, neither to the final expression for the widths nor to the equation for $\eta$.  
A general solution to~\eqref{etaEqij} can be expressed as a superposition of an antisymmetric and a symmetric solution,
\be
\eta(u)=c_a\eta_a(u)+c_s\eta_s(u) \;,
\ee
where
\be
\eta_a(0)=0 \quad \eta_a'(0)=1 
\quad
\eta_s(0)=1 \quad \eta_s'(0)=0 \;.
\ee
For $j\ne3$, we have from~\eqref{initialDeltaij} $\delta t_{(j)}'(\pm\infty)=\delta z_{(j)}'(\pm\infty)=0$, which means $\eta_{(j)}'(\pm\infty)=0$. Since $\eta_{a,s}'(\infty)\ne0$, this implies $\eta_{(j)}(u)=0$. So only $\eta_{(3)}$ is nonzero. From~\eqref{initialDeltaij} we have $\eta_{(3)}'(\pm\infty)=-1/(2p_0)$ and hence
\be
\eta_{(3)}(u)=-\frac{1}{2p_0}\frac{\eta_a(u)}{\eta_a'(\infty)} \;.
\ee
Substituting into~\eqref{ADeltaij} gives
\be
d_{\Delta,z}^{-2}=\mathcal A^\Delta_{33}=\frac{1}{2p_0^2} \Im \left( \frac{t}{p_0}-\frac{\eta_a}{\eta_a'}\right)(\infty) \;.
\ee
Thus, the off-diagonal components of $\mathcal A^\Delta_{ij}$ are zero.

Next we perform the $P_j$ integrals following essentially the same steps. For the first derivative we have
\be\label{eq:MomentumSaddle}
\frac{\pa \mathcal A}{\pa P_i} = 4\Im \left[ \frac{P_i}{p_0} t - q^i\right](\infty) \;.
\ee
Setting $\frac{\pa \mathcal A}{\pa P_i}=0$ determines the saddle point for $P_i$. We again only need the first-order variation of the instanton with respect to $\delta P_j=P_j-P_{sj}$,
\be
q^\mu \to q^\mu + \delta P_j \, \delta q^\mu_{(j)}+\mathcal{O}(\delta P^2) \;.
\ee
The equation for $\delta q^\mu_{(j)}$ is the same as before~\eqref{deltaqeq}, but the asymptotic boundary conditions are different,
\be
\delta q^{i\prime}_{(j)}(\pm\infty)=\delta_{ij} 
\qquad
\delta t'_{(j)}(\pm\infty) =\pm\frac{P_j}{p_0} \; ,
\ee
which follows from expanding $q_j'(\pm\infty)=-P_j$. We find
\be
\begin{split}
&\mathcal A^P_{ij} :=\frac{1}{2}\frac{\pa^2 \mathcal A}{\pa P_i \, \pa P_j} \\
&= 2\Im \left[ -\delta q^i_{(j)}(u_1) + \delta t_{(j)} \frac{P_i}{p_0} + \frac{t}{p_0}\left( \delta_{ij} - \frac{P_i P_j}{p_0^2} \right) \right](\infty) \;.
\end{split}
\ee
The off-diagonal terms vanish as before, and
\be
\delta x_{(1)}(u)=\frac{\delta x_a(u)}{\delta x_a'(\infty)}
\qquad
\eta_{(3)}(u)=\frac{1}{p_0}\frac{\eta_s(u)}{\eta_s'(\infty)} \;,
\ee
which gives
\be
\begin{split}
d_{P,\perp}^{-2} &= \mathcal A^P_{11}= \mathcal A^P_{22} =2 \Im \left(\frac{t}{p_0} -\frac{\delta x_a}{\delta x_a'}\right)(\infty) \\
d_{P,z}^{-2} &= \mathcal A^P_{33} =\frac{2}{p_0^2} \Im \left(\frac{t}{p_0} -\frac{\eta_s}{\eta_s'}\right)(\infty) \;.
\end{split}
\ee

Thus, we have four independent widths,
\be\label{eq:Widths}
\begin{split}
d_{\Delta,z}^{-2}&=\frac{1}{2p_0^2} \Im \left( \frac{t}{p_0}-\frac{\eta_a}{\eta_a'}\right) 
\quad
d_{P,z}^{-2}=\frac{2}{p_0^2} \Im \left(\frac{t}{p_0} -\frac{\eta_s}{\eta_s'}\right)\\
d_{\Delta,\perp}^{-2}&=\frac{1}{2} \Im \left(\frac{t}{p_0} -\frac{\delta x_s}{\delta x_s'}\right) 
\quad
d_{P,\perp}^{-2}=2 \Im \left(\frac{t}{p_0} -\frac{\delta x_a}{\delta x_a'}\right) \;,
\end{split}
\ee
where all quantities are evaluated at $u=\infty$. Note that, apart from the instanton, the widths are obtained from solutions to~\eqref{deltaxj} and~\eqref{etaEqij} which have simple initial conditions at $u=0$. In other words, there is no need to use a shooting method for these additional functions. 

Choosing the contour such that $\text{Im }t=0$ for $r>r_c$,  
\be\label{dLocalNonlocal}
\begin{split}
d_{\Delta,z}^{-2}&=\frac{1}{2p_0^2}\text{Im}\left(-\frac{\eta_a}{\eta_a'}\right)=\frac{W(\eta_{ar},\eta_{ai})}{2p_0^2|\eta_a'|^2} \\
d_{P,z}^{-2}&=\frac{2}{p_0^2}\text{Im}\left(-\frac{\eta_s}{\eta_s'}\right)=2\frac{W(\eta_{sr},\eta_{si})}{p_0^2|\eta_s'|^2} \\
d_{\Delta,\perp}^{-2}&=\frac{1}{2}\text{Im}\left(-\frac{\delta x_s}{\delta x_s'}\right)=\frac{W(\delta x_{sr},\delta x_{si})}{2|\delta x_s'|^2} \\
d_{P,\perp}^{-2}&=2\text{Im}\left(-\frac{\delta x_a}{\delta x_a'}\right)=2\frac{W(\delta x_{ar},\delta x_{ai})}{|\delta x_a'|^2} \;,
\end{split}
\ee
where $W(f,g)=fg'-f'g$ is the Wronskian, $\eta_{ar}=\text{Re }\eta_a$ and $\eta_{ai}=\text{Im }\eta_a$ etc.

$h$ is the same as in~\cite{DegliEsposti:2022yqw}, but we can simplify it further using the above ideas. We start with Eq.~(130) in~\cite{DegliEsposti:2022yqw}, but rewrite it in terms of the normalized solutions~\eqref{initialConditionsetadelta} as (note that we used different notation in~\cite{DegliEsposti:2022yqw})  
\be
|h|=2\left|\frac{\eta_s}{\eta_s'}-\frac{\eta_a}{\eta_a'}\right|^{-1} \;.
\ee
Since the Wronskian of $\eta_s$ and $\eta_a$ is constant (for all $u$), we have $(\eta_s\eta_a'-\eta_a\eta_s')(u)=(\eta_s\eta_a'-\eta_a\eta_s')(0)=1$ and hence
\be
|h|=2|\eta_s'\eta_a'| \;.
\ee

We can obtain a similar expression for $\bar{\phi}$.
We first note that $\bar{\phi}$ satisfies the same equation as $\delta x$, so we can write $\bar{\phi}=c_a\delta x_a+c_s\delta x_s$, where $c_a$ and $c_s$ are two constants that we determine using the initial conditions~\eqref{initialbarphi} and~\eqref{initialConditionsetadelta}. We find
\be
|\bar{\phi}'(u_1)|=\frac{2|\delta x_s'\delta x_a'|}{|\delta x_s\delta x_a'-\delta x_a\delta x_s'|}=2|\delta x_s'\delta x_a'| \;,
\ee
where in the second step we have used the fact that Wronskian of $\delta x_s$ and $\delta x_a$ is constant and evaluated it at $r=0$.

\section{Instantons on the complex plane}\label{Complex plot}

In the main text we argue that the most convenient contour for this class of fields, especially for $\gamma\ll1$, is a path travelling along the imaginary axis from the origin to an the imaginary value $u_c$, then parallel to the real axis towards infinity. Although this single contour is sufficient to compute the full spectrum, it is interesting to consider the instantons as complex-variable functions. To obtain such functions, we have to numerically solve the Lorentz-force equation along a large set of contours starting from $u = 0$ (after we have found the turning point $t(0)$). 

Since we expect singularities along the real axis and a periodic structure along the imaginary axis, one possible choice can be the following: we start with a single contour along the imaginary axis $u_i(r) = ir$ and obtain solutions $t_i(r) := t(ir)$, $z_i(r) := z(ir)$. Then, these functions act as a set of initial conditions which we use to solve parallel to the real axis along a set of contours $u_R(r) = iR + r$ for several values of $R$, obtaining solutions $t_R(r) = t(iR+r)$ and $z_R(r) = z(iR+r)$. Solving for a function effectively of two variables (real/imaginary parts of $u$) using initial conditions at a single point is possible only because the solutions are analytic everywhere except at the branch points.

\begin{figure}[!ht]
\centering
\includegraphics[width=\linewidth]{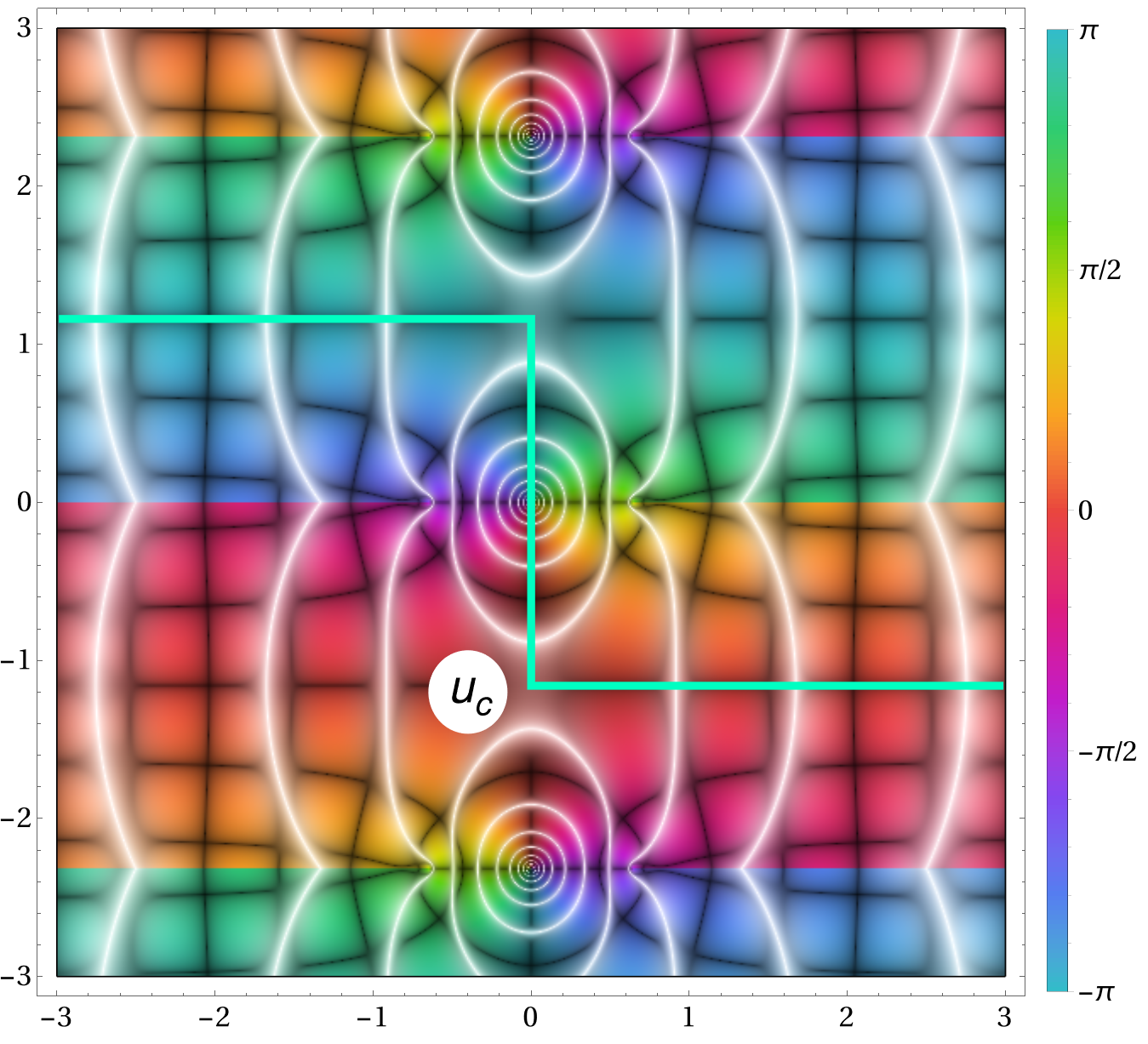}
\caption{$z(u)$ on the complex $u$ plane for $\gamma=1$. We see that along the physical contour $z(u)$ is always real.}
\label{fig:zcomplex}
\end{figure}

In order to visualize the resulting functions there are several possibilities. Since we are mostly interested in the phase, we color the complex $u$ plane depending on the phase of $q(u)$ and add lines of constant real/imaginary part of $q$. The result is shown in the main text in Fig.~\ref{fig:instanton2} for the $t$ component and in Fig.~\ref{fig:zcomplex} for $z$. We see in particular that, since at $u_c$ both the real and imaginary part are zero and constant along black lines, $t(u)$ is either purely real or imaginary along the ``physical'' contour.

\begin{figure}[!ht]
\centering
\includegraphics[width=\linewidth]{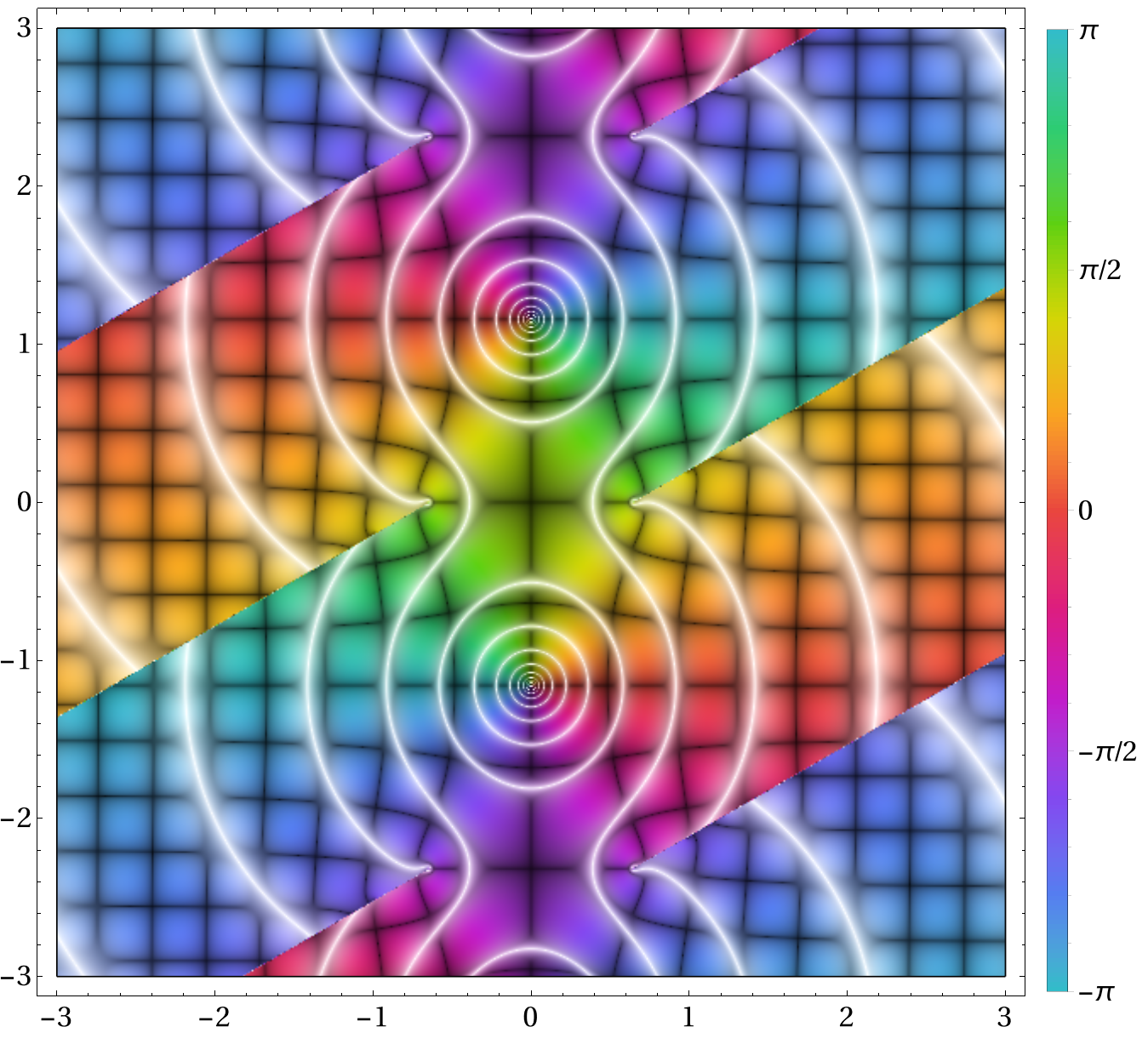}
\caption{$t(u)$ on the complex $u$ plane for $\gamma=1$
with rotated branch cuts. The angle of the cuts is $\theta_c = \frac{\pi}{6}$.}
\label{fig:rotated}
\end{figure}

Functions of a complex variable can have branch points. If the area enclosed by two paths from the origin to some value $u$ contains a branch point, the value $q(u)$ will be different even if it is analytic. 
In fact, Fig.~\ref{fig:instanton2} shows that there is a periodic set of branch points, with cuts parallel to the real line due to our choice of contours. If we rotate the contours $u_R(r)$ by some phase we obtain rotated branch cuts as in Fig.~\ref{fig:rotated}, allowing us to see a different Riemann sheet. The existence of such branch points is directly related to singularities of the field. Since the initial conditions are imaginary and $E(z,t)$ is real when $z$ and $t$ are imaginary, both $t$ and $z$ will continue to be imaginary when $u$ follows the real axis. For the pulse shapes we consider, $g(t)$ either diverges at $t\to i\infty$ or hits a pole at a finite $t=i|t_p|$. In both cases the instantons will cross a singularity of the field if the $u$ contour is along the real axis. However, the situation is qualitatively different for a Gaussian pulse and for a Lorentzian/Sauter pulse. While the first has an essential singularity at infinity, which makes the instantons divergent at branch points, the other two have poles along the imaginary axis, so the instantons remain finite. One can see this already in the simpler time dependent case. Let $E(t)$ be be a field with a pole of order $\beta$ at $t_p$ and expand the instantons around the branch point $u_B$ with an ansatz
\be
E(t) \sim \frac{R}{(t - t_p)^\beta}, \qquad t(u) \sim t_p + c_t(u-u_B)^\alpha
\ee
and similarly for $z$. Plugging this into the Lorentz force equation we see that $\alpha = 1/\beta$, therefore for a field like a Sauter pulse with a double pole the branch point is like a square root $t(u) \sim t_p + c_t\sqrt{u-u_B}$. This method does not give the correct result for a field with a simple pole like a Lorentz pulse, indicating that near the branch point the instanton is not approximated by $(u-u_B)^\alpha$ for any fractional power $\alpha$. This is related to the fact that $A(t)$ itself has a branch point of log-type when $A'(t) = E(t)$ has a simple pole. On the other hand, one also sees that for the Gaussian pulse we have $t(u) \sim \sqrt{\ln(u-u_B)}$.
Due do Liouville's theorem, we always have singularities except for constant fields. Indeed the constant field instantons~\eqref{t0z0} are trivially entire functions.

Furthermore, for a field with poles, since the field is given by a dimensionless function $f(v)$ with a pole $v_p$ and $v = \omega t$, as $\omega$ grows, the pole $t_p$ moves closer to the origin. Since the turning point is squeezed between the origin and the pole, it will get closer to the latter. From this it also follows that the branch cuts move closer to the origin. This makes it numerically more challenging to reach larger $\omega$ values for such fields. 

\begin{figure}[!ht]
\centering
\includegraphics[width=\linewidth]{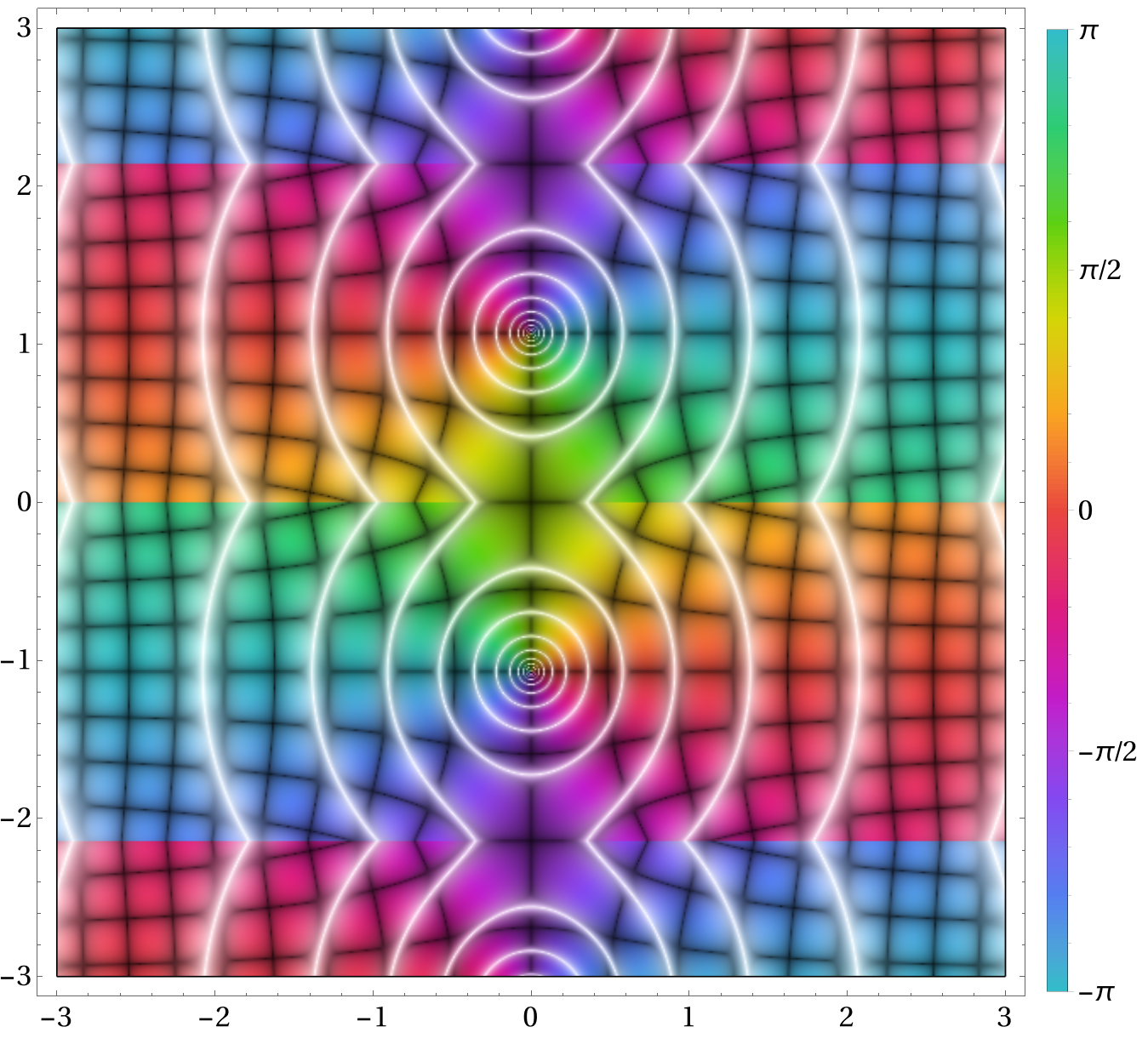}
\caption{$t(u)$ on the complex $u$ plane for $\gamma=1$
for the Lorentz pulse.}
\label{fig:tLorentz}
\end{figure}

\begin{figure}[!ht]
\centering
\includegraphics[width=\linewidth]{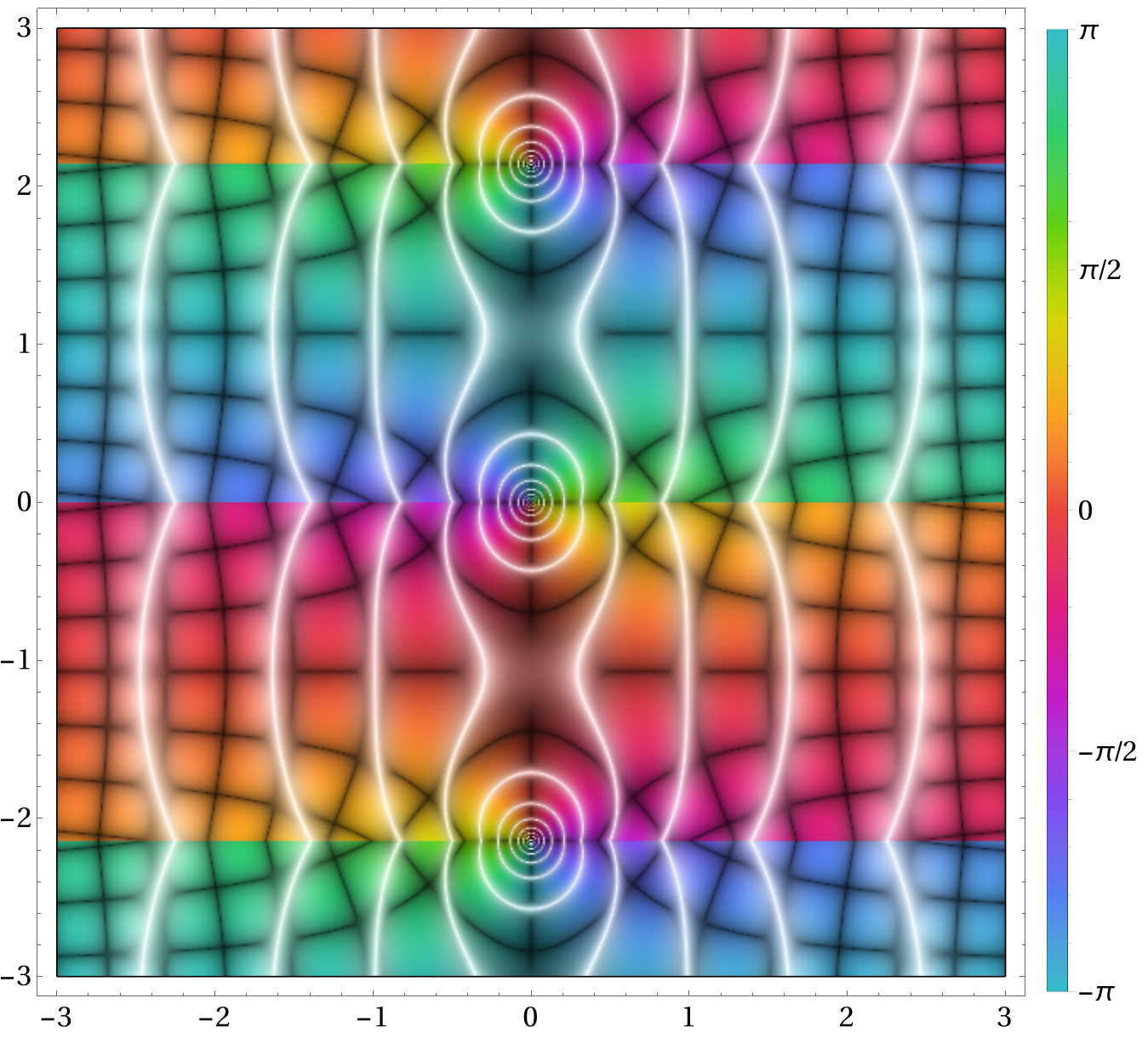}
\caption{$z(u)$ on the complex $u$ plane for $\gamma=1$
for the Lorentz pulse. Both components look very similar to the solutions for a Gaussian pulse. The main difference is the behavior near the branch points.}
\label{fig:zLorentz}
\end{figure}

\section{Additional plots}

\begin{figure}[!ht]
\centering
\includegraphics[width=\linewidth]{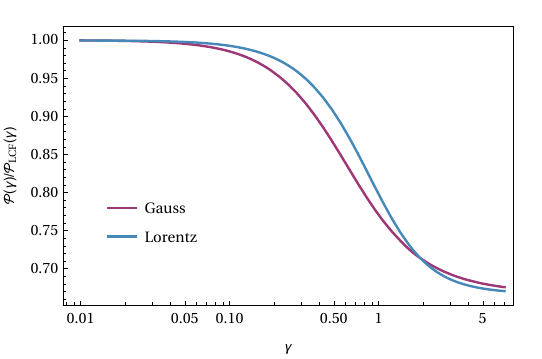}
\caption{Saddle point value of the longitudinal momentum as a function of $\gamma$ normalized by the corresponding analytical expression of the $\gamma \to 0$ limits, namely $\frac{3\sqrt{\pi}}{4 \gamma}$ for the Gaussian pulse and $\frac{3 \pi}{4 \gamma}$ for the Lorentzian pulse.}
\label{fig:Psaddle}
\end{figure}

\begin{figure}[!ht]
\centering
\includegraphics[width=\linewidth]{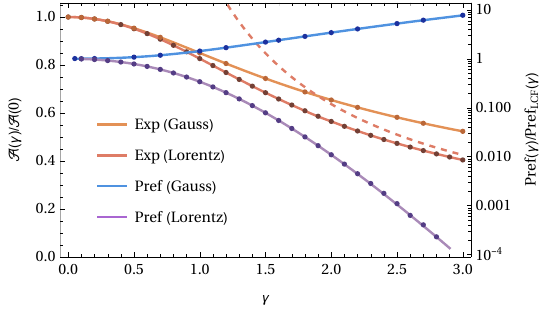}
\caption{Exponent and prefactor for the Gaussian and Lorentzian pulses and comparison with the effective action (dots). The action is qualitatively similar for the two fields, but for the Lorentz pulse it approaches the leading-order perturbative result~\eqref{perturbativeLorentz} (dashed line) at large $\gamma$. On the other hand, the prefactors behave very differently at larger values of $\gamma$.}
\label{fig:overlapped}
\end{figure}

\begin{figure}[!ht]
\centering
\includegraphics[width=\linewidth]{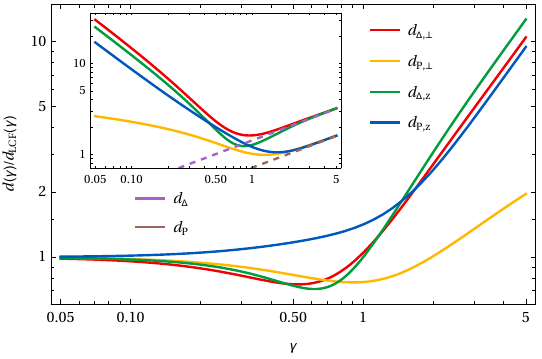}
\caption{All four widths for the Lorentz pulse. We can see that qualitatively they look similar to Fig.~\ref{fig:comparison} for the Gaussian pulse. At large $\gamma$ we find agreement with~\eqref{dlargeGammaApprox} (dashed lines).}
\label{fig:widthsLorentz}
\end{figure}

In the main text we show the result for the exponent, the prefactor and the widths for the Gaussian pulse, $g'''(t)=e^{-(\omega t)^2}$, but since the analytical results are valid for a general pulse shape, we considered also a Lorentzian pulse, $g'''(t)=1/(1+[\omega t]^2)$, and compared the two. In Figs.~\ref{fig:tLorentz} and~\ref{fig:zLorentz} we show $t(u)$ and $z(u)$ in the complex $u$ plane. Although the Lorentzian has a pole, these complex plots look quite similar to Figs.~\ref{fig:instanton2} and Fig.~\ref{fig:zcomplex} for the Gaussian field.   
In Fig~\ref{fig:Psaddle} we see the maximum of the longitudinal momentum for both field shapes, normalized by their $\gamma \to 0$ limits $H(\infty)/\gamma$ from Appendix~\ref{Dipole fields}. In Fig.~\ref{fig:overlapped} we see the exponent and prefactor for both fields and their agreement with the effective action. We comment on the qualitative difference between the prefactors in Appendix~\ref{LCF expansions in the formation region}. In Fig.~\ref{fig:widthsLorentz} we see all four widths for the Lorentzian pulse normalized by their LCF results.

\section{LCF expansions in the formation region}\label{LCF expansions in the formation region}

In the formation region $t$ and $z$ are not large, so we can expand the field in~\eqref{Fdef} as
\be\label{Fgamma2}
F(t,z)\approx G^{(3)}(0)+\frac{G^{(5)}(0)}{2}\left(t^2+\frac{z^2}{5}\right)\gamma^2 \;,
\ee
where $G^{(3)}=G'''$ etc. We set
\be\label{G30G50}
G^{(3)}(0)=1
\qquad
G^{(5)}(0)=-2 \;,
\ee
where the first condition means $E$ is the maximum field strength, and the second is used to define $\omega$. There is no loss of generality in these choices for $G^{(3)}(0)$ and $G^{(5)}(0)$. They just define what we mean by $E$ and $\omega$. For example, $\exp(-[\omega t]^2)$ and $\exp(-[2\omega t]^2)$ are the same functions, just with different normalization of $\omega$ or $G^{(5)}(0)$. However, the relative factor of $5$ between the $t^2$ and $z^2$ terms cannot be changed. It just happens to be this factor for all e-dipole fields.
We chose $G^{(5)}(0)=-2$ so that the coefficient of $t^2$ is simple, which means $E_3(t,z=0)=Eg'''(t)$ is simple. For $\gamma\ll1$ one might instead want to choose a simple $E_3(t=z)$, which would mean a different $G^{(5)}(0)$ would be simpler.

We solve the Lorentz force equation with the ansatz $t\approx t_0(u)+t_1(u)\gamma^2$ and $z\approx z_0(u)+z_1(u)\gamma^2$. To leading order we find 
\be\label{t0z0}
t_0(u)= i\cosh u
\qquad
z_0(u)= i\sinh u \;.
\ee
For the next order we use initial conditions $z_1(0)=z_1'(0)=t_1'(0)=0$, while $t_1(0)$ is a constant to be determined. The $u$ contour starts at $u=0$ and follows the negative imaginary axis. Near $u=-i\pi/2$ the contour turns and goes parallel to the real axis\footnote{For the numerical solution without using LCF, we choose a contour with a smooth turn.}. We use $u_c$ to refer to the exact point where the contour turns and where $t$ becomes real. We have $u_c\approx-\frac{i\pi}{2}+\delta u\gamma^2$.
We determine the two constants, $t_1(0)$ and $\delta u$, by demanding that $t(u_c)=0$ and $z'(u_c)=0$. 
We find 
\be
t_1(0)=-\frac{i}{5}
\qquad
\delta u=\frac{i\pi}{5}
\ee
and
\be
\begin{split}
t_1(u)&=\frac{i}{20}[8u\sinh(u)-5\cosh(u)+\cosh(3u)] \\
z_1(u)&=\frac{i}{20}[8u\cosh(u)-11\sinh(u)+\sinh(3u)] \;.
\end{split}
\ee

For the longitudinal widths we need 
\be
\eta_{a}^{(0)}(u)=\sinh(u)
\qquad
\eta_{s}^{(0)}(u)=\cosh(u)
\ee
and
\be
\begin{split}
\eta_a^{(1)}(u)&=\frac{1}{20}[4u\cosh(u)-13\sinh(u)+3\sinh(3u)] \\
\eta_s^{(1)}(u)&=\frac{1}{5}\sinh(u)[7u+3\cosh(u)\sinh(u)] \;.
\end{split}
\ee
Evaluating these at $u=u_c$ gives us the Wronskians in~\eqref{dLocalNonlocal}
\be\label{Wlead}
\eta_{ar}\eta_{ai}'-\eta_{ai}\eta_{ar}'\approx\frac{\pi}{10}\gamma^2 
\qquad
\eta_{sr}\eta_{si}'-\eta_{si}\eta_{sr}'\approx\frac{\pi}{2}\gamma^2 \;.
\ee

For the transverse widths we need 
\be
\delta x_a^{(0)}=u
\qquad
\delta x_s^{(0)}=1
\ee
and
\be
\begin{split}
\delta x_a^{(1)}&=\frac{3}{20}\left[\sinh(2u)-u\cosh(2u)-u-\frac{4}{9}u^3\right] \\
\delta x_s^{(1)}&=\frac{1}{20}\left[-3\cosh(2u)+3-4u^2\right] \;.
\end{split}
\ee
Evaluating these at $u_c$ gives
\be\label{deltaxWronskiansLCF}
\delta x_{sr}\delta x_{si}'-\delta x_{si}\delta x_{sr}'\approx\frac{\pi}{5}\gamma^2 
\quad
\delta x_{ar}\delta x_{ai}'-\delta x_{ai}\delta x_{ar}'\approx\frac{\pi}{2} \;.
\ee

The above results give the LO contribution from the formation region, which we will combine with the LO contribution from the acceleration region in Appendices~\ref{The longitudinal widths} and~\ref{The transverse widths} to obtain the widths to LO. However, to explain the qualitatively different prefactors for the Gaussian and the Lorentzian pulses seen in Fig.~\ref{fig:overlapped}, we have to consider at least the NLO contribution from the formation region (recall that the acceleration region does not contribute to the prefactor). 

We obtain the NLO in the same way as above, i.e. by just expanding each quantity to one power higher in $\gamma^2$, e.g. $q\approx q^{(0)}+q^{(1)}\gamma^2+q^{(2)}\gamma^4$. 
$q^{(2)}$, $\eta^{(2)}$ and $\delta x^{(2)}$, can again be expressed in terms of powers of $u$, and $\cosh$ and $\sinh$, but the expressions are not particularly illuminating. For the $u$ independent quantities we find
\be
u_c\approx i\left(-\frac{\pi}{2}+\frac{\pi}{5}\gamma^2+\frac{3\pi}{560}[G^{(7)}(0)-28] \gamma^4 \right)
\ee
\be
t(0)\approx i\left(1-\frac{1}{5}\gamma^2+\left[\frac{1}{75}-\frac{G^{(7)}(0)}{280}\right]\gamma^4\right)
\ee
and
\be
\begin{split}
W(\eta_{ar},\eta_{ai})&\approx\frac{\pi}{10}\gamma^2+\frac{\pi}{280}\left[G^{(7)}(0)-70\right]\gamma^4 \\
W(\eta_{sr},\eta_{si})&\approx\frac{\pi}{2}\gamma^2+\frac{\pi}{40}[G^{(7)}(0)-6]\gamma^4 \\
W(\delta x_{ar},\delta x_{ai})&\approx\frac{\pi}{2}+\frac{\pi}{60}[2\pi^2-21]\gamma^2 \\
W(\delta x_{sr},\delta x_{si})&\approx\frac{\pi}{5}\gamma^2\\
&+\frac{\pi}{8400}[90G^{(7)}(0)+112\pi^2-1029]\gamma^4 \;,
\end{split}
\ee
where $G^{(7)}(0)=\partial_x^7 G(x)|_{x=0}$. Since the field is assumed to be symmetric, $G^{(7)}(0)$ is the first nonzero derivative that is not fixed by the normalization of the field strength and $\omega$.  
Inserting this into the prefactor part of~\eqref{PrefWronskians} gives
\be\label{PrefNLO}
\begin{split}
\text{Pref}&\approx\frac{5\sqrt{5}}{(2\pi)^3\gamma^4}\bigg[1+\frac{4557-224\pi^2-162G^{(7)}(0)}{1680}\gamma^2\bigg] \\
&\approx\frac{5\sqrt{5}}{(2\pi)^3\gamma^4}(1+[1.4-0.096G^{(7)}(0)]\gamma^2) \;.
\end{split}
\ee
Thus, as $\gamma$ increases, the ratio of the prefactor and its leading-order approximation, $\text{Pref}/\text{Pref}_{LO}$, becomes either larger or smaller depending on whether $G^{(7)}(0)$ is smaller or larger than
\be
\frac{4557-224\pi^2}{162}\approx14.5 \;.
\ee
For a Gaussian pulse, $G'''(x)=e^{-x^2}$, we have $G^{(7)}(0)=12$ and
\be
\frac{\text{Pref}}{\text{Pref}_{LO}}\approx1+0.24\gamma^2 \;,
\ee
while for a Lorentzian pulse, $G'''(x)=1/(1+x^2)$, we have $G^{(7)}(0)=24$ and
\be
\frac{\text{Pref}}{\text{Pref}_{LO}}\approx1-0.92\gamma^2 \;.
\ee
This explains the qualitatively different prefactors seen in Fig.~\ref{fig:overlapped}.

\begin{figure}[!ht]
\centering
\includegraphics[width=\linewidth]{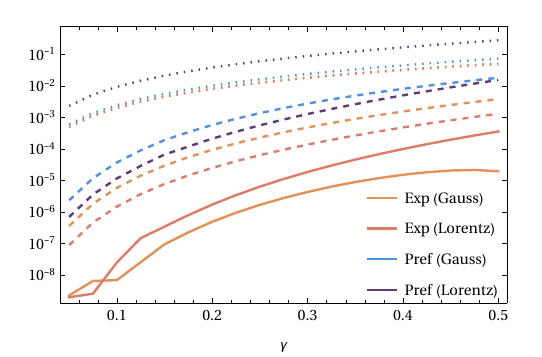}
\caption{Relative error of first orders in the $\gamma\ll1$ expansion of the exponent~\eqref{AedipoleNLO} and the prefactor~\eqref{PrefNLO}, with dotted lines for the leading order, the dashed lines for LO$+$NLO, and solid lines for LO$+$NLO$+$NNLO.}
\label{fig:LCF}
\end{figure}

In Fig~\ref{fig:LCF} we see a comparison of the action and the prefactor with their expansions. We plot
\be
\Delta \mathcal A := \frac{\mathcal A_{\text{approx}}}{\mathcal A_{\text{exact}}} - 1 \; ,
\ee
with $A_{\text{approx}}$ representing the expansion up to LO (dotted), NLO (dashed), and NNLO (solid), and similarly for the prefactor. We see that by including these first couple of terms we obtain a good approximation all the way up to $\gamma\sim0.5$, which is not particularly small. The noisy error seen in Fig~\ref{fig:LCF} around $\gamma\sim0.1$ for NNLO for the exponent is due to the numerical precision rather than the error of the analytical approximation.    

Inserting the $\gamma\ll1$ expansions just found into~\eqref{Adef} and expanding the field gives
\be\label{AedipoleNLO}
\mathcal{A}\approx\frac{\pi}{E}\left(1-\frac{\gamma^2}{5}-[G^{(7)}(0)-28]\frac{\gamma^4}{280}\right) \;.
\ee
Increasing $\gamma$ thus leads to a reduction of the exponential suppression and therefore to a larger probability. The same happens for a purely time dependent electric field, while the opposite happens for a purely $z$ dependent field.

We can generalize the e-dipole result~\eqref{AedipoleNLO} to a general field, i.e. we calculate the NLO correction in
\be
\mathcal{A}(\gamma)\approx\mathcal{A}(0)+\frac{1}{2}\mathcal{A}''(0)\gamma^2 \;.
\ee
We begin by writing 
\be
\mathcal A(\gamma) = 2\Im \left[p x_\LCp +p'x_\LCm -\frac{T}{2} -\int_0^1\ud\tau\left(\frac{\dot{q}^2}{2T}+A\dot{q}\right) \right] \;.
\ee
Since all the integration variables are evaluated at their saddle-point values, the total $\gamma$ derivative is equal to
\be
\frac{1}{2}\mathcal{A}''(0)=-\lim_{\gamma\to0}\frac{1}{\gamma}\int_{-\infty}^\infty\ud u\frac{\ud A^\mu}{\ud\gamma}q'_\mu \;.
\ee
The derivative with respect to $\gamma$ is up to a factor of $E$ equal to the derivative with respect to the frequency, and is therefore not affected by our rescaling $q^\mu\to q^\mu/E$ and $u\to u/E$. We can express the $\gamma$ dependence of the field as $A_\mu(q)=f_\mu(\gamma q)/\gamma$. To take the $\gamma\to0$ limit we need to expand $f_\mu(\gamma q)$ up to $\mathcal{O}(\gamma^3)$. Even though this is the NLO correction to the exponent, we only need the zeroth order approximation of the instanton, $q\approx q_{(0)}$, given by~\eqref{t0z0}, and $u_c\approx-i\pi/2$. Only the part of the $u$ contour from $+i\pi/2$ to $-i\pi/2$ contributes to the imaginary part. We have
\be
\frac{1}{2}\mathcal{A}''(0)=-\frac{1}{3E}\text{Im}\int_{i\pi/2}^{-i\pi/2}\ud u f_{\mu,\nu\rho\sigma}q^{\prime\mu}q^\nu q^\rho q^\sigma \;.
\ee
Substituting~\eqref{t0z0} for $q$ gives elementary integrals. We find
\be\label{AgenNLO}
\mathcal{A}\approx\frac{\pi}{E}\left(1+\frac{\gamma^2}{8}[F_{00}(0)-F_{33}(0)]\right) \;,
\ee
where, in terms of the usual $t$ and $z$ (not rescaled by $E$), $F(\omega t,\omega z)=E_3(t,z)$, $F_{00}(0)=\partial_{\omega t}^2 E_3(t=0,z=0)/E$ and $F_{33}(0)=\partial_{\omega z}^2 E_3(t=0,z=0)/E$. For example, for an e-dipole field we have $F_{00}(0)=-2$ and $F_{33}=-2/5$ from~\eqref{Fgamma2}, and we recover~\eqref{AedipoleNLO}. 

For a purely time-dependent Sauter pulse, $E_3(t)=E\text{sech}^2(\omega t)$, we have $F_{00}(0)=-2$ and $F_{33}=0$, and~\eqref{AgenNLO} gives
\be\label{ALCFforEoft}
\mathcal{A}\approx\frac{\pi}{E}\left(1-\frac{\gamma^2}{4}\right) \;,
\ee
which agrees with the expansion of the exact result~\cite{Popov:1972,Dunne:2005sx,Dunne:2006st} for $\mathcal{A}$,
\be
\mathcal{A}=\frac{\pi}{E}\frac{2}{1+\sqrt{1+\gamma^2}} \;.
\ee

A purely $z$ dependent field, e.g. a Sauter pulse $E_3(z)=E\text{sech}^2(\omega z)$, would lead to the same correction but with opposite sign. This is expected. Increasing (decreasing) $\gamma$ for a time ($z$) dependent field leads in general to a larger (smaller) probability. Since the correction in~\eqref{AedipoleNLO} is negative, an e-dipole field behaves more like a time-dependent field.

Note that, while we only needed $q^\mu_{(0)}(u)$, which also gives the instanton for a constant field, the result~\eqref{AgenNLO} cannot be obtained from the standard LCF approximation~\eqref{standardLCF}. Note also that the correction can be numerically important, because while $\gamma^2\ll1$, $\gamma^2/E$ is not necessarily small.

\section{The longitudinal widths}\label{The longitudinal widths}

In the previous section we calculated the local parts of the LCF approximation. Now we turn to the nonlocal parts, which are more challenging.

As explained in the main text, to leading order we have 
\be
\phi''_0=F(\phi_0)\phi_0' \;.
\ee
With initial conditions $\phi_0(0)=\phi_0'(0)=i\gamma/2$, the solution is
\be\label{dphiIntF}
\phi_0'(u)=\frac{i\gamma}{2}+\int_{i\gamma/2}^{\phi_0(u)}\!\ud\varphi\, F(\varphi) 
=H(\phi_0)+\mathcal{O}(\gamma^3) \;.
\ee

For the other lightfront variable, we have a first-order equation $\theta'=\frac{\gamma^2}{2\phi'}$ and (approximate) initial condition $\theta(0)=i\gamma$, so the solution is given by  
\be\label{thetaA}
\theta(u)=i\gamma+\frac{\gamma^2}{2}\int_0^u\frac{\ud v}{\phi'(v)} \;.
\ee

The correction to $\phi\approx\phi_0+\delta\phi$ is determined by
\be
\delta\phi''=F(\phi_0)\delta\phi'+[F'(\phi_0)\delta\phi+F_\theta(\phi_0)\theta]\phi_0' \;,
\ee
where 
\be\label{FthetaDef}
F_\theta(\phi)=\partial_\theta F(\phi,\theta=0) \;.
\ee
But it turns out that we actually do not need $\delta\phi$. 
To keep the notation simple, from now on we will write $\phi$ instead of $\phi_0$.

For $\eta$ we have $\eta\approx\eta_0$, where
\be\label{eta0eq}
\eta_0''=[F^2(\phi)+F'(\phi)\phi']\eta_0 \;.
\ee
One solution to this equation is $\eta_0=\phi'$. A second independent solution can be obtained using Abel's identity, allowing us to write a general solution as
\be\label{eta0gensol}
\eta_0(u)=\phi'(u)\left(a+b\int_0^u\frac{\ud v}{\phi^{\prime2}(v)}\right) \;,
\ee
where $a$ and $b$ are two constants. Imposing the initial conditions~\eqref{initialConditionsetadelta} we find
\be
\eta_{a0}(u)=\frac{i\gamma}{2}\phi'(u)\int_0^u\frac{\ud v}{\phi^{\prime2}(v)}
\ee
and
\be
\eta_{s0}(u)=\phi'(u)\left(\frac{2}{i\gamma}-\frac{i\gamma}{2}F\left[\frac{i\gamma}{2}\right]\int_0^u\frac{\ud v}{\phi^{\prime2}(v)}\right) \;,
\ee
where we can approximate $F(i\gamma/2)\approx1$. Close to $u=0$ we have $\phi'=\mathcal{O}(\gamma)$, so there $\eta_0=\mathcal{O}(1)$. Outside the formation region, as $\phi'$ becomes $\mathcal{O}(1)$, we have $\eta_0=\mathcal{O}(1/\gamma)$. Asymptotically we have
\be\label{eta0derAsymp}
\eta_{a0}'(\infty)=\frac{i\gamma}{2\phi'(\infty)}\approx-\eta_{s0}'(\infty) \;.
\ee
Since $\phi'(\infty)=\mathcal{O}(1)$, we have $\eta_{a0}'(\infty),\eta_{s0}'(\infty)=\mathcal{O}(\gamma)$. Thus, in both cases there are regions where $\eta_0$ is one order of magnitude larger than the asymptotic $\eta_0'$. As we will now show, the ``next-order'' correction to~\eqref{eta0eq} will actually contribute to the same order of magnitude for $\eta'(\infty)$. 

The equation for the next-order is
\be\label{deltaetaeq}
\delta\eta''=[F^2(\phi)+F'(\phi)\phi']\delta\eta+R\eta_0 \;,
\ee
where $R$ is a function of $\phi$, $\theta$ and $\delta\phi$. By separating out a factor of $\phi'$ as
\be
\delta\eta(u)=\phi'(u)\varepsilon(u)
\ee
we obtain a simpler equation for $\varepsilon(u)$,
\be
\varepsilon''(u)+2F\varepsilon'(u)=R\frac{\eta_0}{\phi'} \;.
\ee
We can solve this equation using $F(\phi)=\phi''/\phi'$,
\be\label{dvarepsilon}
\varepsilon'(u)=\frac{1}{\phi^{\prime2}(u)}\int_0^u\ud v\,\phi^{\prime2}R\frac{\eta_0}{\phi'} \;.
\ee
Asymptotically we have
\be
\delta\eta'(\infty)=\phi'(\infty)\varepsilon'(\infty)=\frac{1}{\phi'(\infty)}\int_0^\infty\!\ud v\,\phi^{\prime2}R\frac{\eta_0}{\phi'} \;.
\ee

$R=R_\theta+R_{\delta\phi}$ has two terms, one ($R_\theta$) proportional to $\theta$ or $\theta'$, and the other ($\delta\phi$) proportional to $\delta\phi$ or $\delta\phi'$. 
We begin with $R_\theta$,
\be
\begin{split}
R_\theta&=-F_\theta\theta'+(2FF_\theta+\phi'F_\theta')\theta \\
&=\frac{1}{\phi^{\prime2}}\left(\theta\frac{\ud}{\ud u}[\phi^{\prime2}F_\theta]-\theta'\phi^{\prime2}F_\theta\right)\;,
\end{split}
\ee
with $F_\theta$ given by~\eqref{FthetaDef}.
Choosing again $G$ as in~\eqref{simplerg} we have
\be\label{FfromH}
F_\theta(\phi)=\frac{3}{4\phi^4}[-3G(2\phi)+2\phi G'(2\phi)]=\frac{\ud}{\ud\phi}\frac{H(\phi)}{\phi} \;.
\ee
Since $H$ goes to a constant~\eqref{Hinfty}, we have for large $\phi$
\be
F_\theta(\phi)\to-\frac{H(\infty)}{\phi^2} \;,
\ee
so $R=\mathcal{O}(1/u^2)$ asymptotically. This would give $R\eta_0=\mathcal{O}(1/u)$ in~\eqref{deltaetaeq} and hence $\delta\eta'=\mathcal{O}(\ln u)$, which does not agree with the fact that $\delta\eta'$ should go to a constant. This apparent problem is due to the fact that we have expanded $G(\theta)$ and $G'(\theta)$ in $\theta\ll1$. But from~\eqref{thetaA} we have
\be\label{thetaAsymptotic}
\theta\to\frac{\gamma^2}{2\phi'(\infty)}u \;,
\ee
so when $u\gtrsim1/\gamma^2$ we can no longer expand $G(\theta)$. For such large $u$ we have $\phi\gtrsim1/\gamma^2$, and from~\eqref{Fdef} we find $F\approx\mathcal{F}(\theta)/\phi^2=\mathcal{O}(1/\phi^2)=\mathcal{O}(\gamma^4)$, where $\mathcal{F}(\theta)$ is some $\mathcal{O}(1)$ function. $F$ is hence very small for $u\gtrsim1/\gamma^2$ and becomes smaller for larger $u$, and so $\delta\eta'$ will not change significantly for $u\gtrsim1/\gamma^2$. To approximate $\delta\eta'$ we can therefore make an expansion for $\theta\ll1$ as long as we stop at some $u=u_1$ which is large $u_1\gg1$ but still $u_1<1/\gamma^2$ to avoid the region where the expansion in $\theta\ll1$ breaks down. 

Returning to the calculation, the contribution to~\eqref{dvarepsilon} coming from $R_\theta$ is
\be
\begin{split}
\varepsilon_{(\theta)}'(u)=&\frac{1}{\phi^{\prime2}(u)}\int_0^u\ud v\left(\theta\frac{\ud}{\ud v}[\phi^{\prime2}F_\theta]-\theta'\phi^{\prime2}F_\theta\right)\\
&\times\left(a+b\int_0^v\frac{\ud w}{\phi^{\prime2}}\right) \;.
\end{split}
\ee
With a partial integration and $\theta'=\gamma^2/(2\phi')$ we find
\be
\begin{split}
\varepsilon_{(\theta)}'(u)=&\theta F_\theta\left(a+b\int_0^u\frac{\ud w}{\phi^{\prime2}}\right)
-\frac{b}{\phi^{\prime2}(u)}\int_0^u\ud v\,\theta F_\theta \\
&-\frac{\gamma^2}{\phi^{\prime2}(u)}\int_0^u\ud v\phi'F_\theta\left(a+b\int_0^v\frac{\ud w}{\phi^{\prime2}}\right) \;,
\end{split}
\ee
where we have dropped the boundary term at $u=0$ since $a(\theta\phi^{\prime2}F_\theta)|_{u=0}/\phi^{\prime2}(u)=\mathcal{O}(a\gamma^4/\phi^{\prime2}(u))$. Using~\eqref{FfromH} to write $\phi' F_\theta=\ud_u(\phi'/\phi)$ and a second partial integration we find
\be\label{esol}
\begin{split}
\varepsilon_{(\theta)}'(u)=&\theta F_\theta\left(a+b\int_0^u\frac{\ud w}{\phi^{\prime2}}\right)
-\frac{b}{\phi^{\prime2}(u)}\int_0^u\ud v\,\theta F_\theta \\
&-\frac{\gamma^2}{\phi^{\prime2}(u)}\bigg\{a\frac{\phi'}{\phi}\bigg|_0^u+b\int_0^u\frac{\ud v}{\phi^{\prime2}}\left(\frac{\phi'}{\phi}\bigg|_v^u\right)\bigg\} \;.
\end{split}
\ee
By comparing~\eqref{esol} with~\eqref{eta0gensol} we can check that $\delta\eta_{(\theta)}=\phi'\varepsilon$ is indeed smaller than $\eta_0$, which justifies the above treatment. However, the derivative is asymptotically on the same order of magnitude. To show this we take the asymptotic limit,
\be
\varepsilon_{(\theta)}'(\infty)=\frac{\gamma^2}{\phi^{\prime2}(\infty)}\left(a+b\int_0^\infty\frac{\ud u}{\phi'\phi}\right) \;,
\ee
where the main contribution to the above integral comes from the formation region where $\phi\approx\phi'\approx (i\gamma/2)e^u$, so
\be
\varepsilon_{(\theta)}'(\infty)\approx\frac{1}{\phi^{\prime2}(\infty)}(\gamma^2a-2b) \;.
\ee
This gives the same result for both $\eta_a$ ($a=0$ and $b=i\gamma/2$) and $\eta_s$ ($a=2/(i\gamma)$ and $b=-i\gamma/2$), 
\be\label{deltadetaAsymp}
\phi'(\infty)\varepsilon_{(\theta)}'(\infty)=-\frac{i\gamma}{\phi'(\infty)} \;,
\ee
which is indeed on the same order of magnitude as~\eqref{eta0derAsymp}.

We will now show that the part coming from $\delta\phi$ is negligible. We have
\be
\begin{split}
R_\phi&=[\phi'F''+2FF']\delta\phi+F'\delta\phi'\\
&=\frac{1}{\phi^{\prime2}}\frac{\ud}{\ud u}\left[\phi^{\prime2}F'\delta\phi\right] \;,
\end{split}
\ee
so with a partial integration we find
\be
\begin{split}
\varepsilon_{(\phi)}'(u)&=\frac{1}{\phi^{\prime2}(u)}\int_0^u\ud v\,\phi^{\prime2}R_\phi\frac{\eta_0}{\phi'}\\
&\approx\delta\phi F'\left(a+b\int_0^u\frac{\ud v}{\phi^{\prime2}}\right)-\frac{b}{\phi^{\prime2}}\int_0^u\ud v\delta\phi F' \;,
\end{split}
\ee
where we have dropped a negligible boundary term at $u=0$. In the asymptotic limit the first two terms go to zero, while the third is $\mathcal{O}(\gamma\delta\phi)$ which is negligible compared to~\eqref{eta0derAsymp}.

Thus, the dominant contributions come from~\eqref{eta0derAsymp} and~\eqref{deltadetaAsymp},
\be
\eta_a'(\infty)\approx-\frac{i\gamma}{2\phi'(\infty)}
\qquad
\eta_s'(\infty)\approx-\frac{3i\gamma}{2\phi'(\infty)} \;,
\ee
and hence, with $p_0\approx\phi'(\infty)/\gamma$, we finally find some very simple results
\be
p_0^2|\eta_a'(\infty)|^2\approx\frac{1}{4}
\qquad
p_0^2|\eta_s'(\infty)|^2\approx\frac{9}{4}
\qquad
p_0^2|h|\approx\frac{3}{2} \;.
\ee

Interestingly, these LCF approximations of the nonlocal parts of the longitudinal widths do not actually depend on the pulse shape $g$. We can understand this by generalizing the above results beyond e-dipole fields. We consider now either some other 4D fields for which the calculation of the longitudinal widths reduces to a 2D problem in a similar way as for the e-dipole fields, or just a 2D field. We assume that the field can be expanded around the maximum as
\be
E_3(t,z)/E_0\approx1-(t^2+az^2)\gamma^2 \;,
\ee
where $a$ is some constant. For e-dipole fields we have $a=1/5$. The calculation of the local parts is the same as before. The generalization of the Wronskians in~\eqref{Wlead} is given by
\be
W(\eta_{sr},\eta_{si})\approx\frac{\pi\gamma^2}{2}
\qquad
W(\eta_{sr},\eta_{si})\approx\frac{\pi a\gamma^2}{2} \;.
\ee
The calculation of the nonlocal parts is also essentially the same, except that $F_\theta(\phi)$, which is still defined as in~\eqref{FthetaDef}, cannot be expressed as in~\eqref{FfromH}, which only holds for e-dipole fields. We can still go through the same steps by writing $F_\theta(\phi)=:IF_\theta'(\phi)$ and choosing the integration constant such that $IF_\theta(\infty)=0$. We find that the right-hand side of~\eqref{deltadetaAsymp} should be multiplied by
\be\label{Jint}
J=-\int_0^\infty\ud\phi\, F_\theta(\phi) \;.
\ee
Thus, the LCF approximation of the longitudinal widths for a general field is given by
\be\label{dLongLCFgen}
d_{P,z}^{-2}=\frac{\pi\gamma^2}{E}\Big(\frac{1}{2}+J\Big)^{-2}
\qquad
d_{\Delta,z}^{-2}=\frac{\pi a\gamma^2}{4E}\Big(\frac{1}{2}-J\Big)^{-2} \;.
\ee
$J$ gives a nonlocal contribution. For all e-dipole fields we can perform the integral in~\eqref{Jint} using~\eqref{FfromH} to find $J=1$. However, $J\ne1$ in general. For example, if $E_3(t,z)=E_3(z,t)$ then $F(\phi,-\theta)=F(\phi,\theta)$, $F_\theta=0$ and $J=0$.  
For a purely time dependent field we have $F(\phi,\theta)=F(\phi+\theta/2)$ and hence $F_\theta(\phi)=F'(\phi)/2$, so $J=1/2$ and $d_{P,z}^{-2}=\pi\gamma^2/E$, which agrees with~\eqref{B12expansion}.
Thus, the longitudinal widths do in fact depend on the field shape, but there exist entire classes of fields that give the same result. 
We also see that if we replace $E_z(t,z)\to E_z(z,t)$ then $d_{P,z}\leftrightarrow d_{\Delta,z}$, up to a factor of $2$.

\section{The transverse widths}\label{The transverse widths}

Next we turn to the transverse widths. From~\eqref{deltaxeq} we have approximately
\be\label{deltaxeqLL}
\delta x''\approx-\frac{1}{2}\phi'F'(\phi)\delta x \;.
\ee
It turns out that the symmetric solution $\delta x_s$ is simpler to approximate, so we will first solve~\eqref{deltaxeqLL} for $\delta x_s$ and then obtain the antisymmetric solution using Abel's identity (similar to~\eqref{eta0gensol}), which gives
\be\label{dxaFromdxs}
\delta x_a(u)=\delta x_s(u)\int_0^u\frac{\ud v}{\delta x_s^2(v)} \;.
\ee

To solve~\eqref{deltaxeqLL} we change variables from proper time $u$ to lightfront time $\phi$. The velocity $\phi'=\ud\phi/\ud u$ can be expressed in terms of $\phi$ using~\eqref{dphiIntF} and~\eqref{HfromF}, $\phi'\approx H(\phi)$. \eqref{deltaxeqLL} becomes
\be\label{deltaxeqH}
H\delta x''(\phi)+F\delta x'(\phi)=-\frac{1}{2}F'(\phi)\delta x \;,
\ee
where now all primes denote derivatives with respect to $\phi$. We want to find the symmetric solution, which has initial conditions as in~\eqref{initialConditionsetadelta}. \eqref{deltaxeqH} should be solved along some complex $\phi$ contour. If $\delta x_s$ depended on $\gamma$ then we would have started the contour at $\phi=i\gamma/2$. At first sight, it might look like we would actually need to do that, because $H(\phi=i\gamma/2)\approx i\gamma/2$, so $\delta x''$ is multiplied by a function that is $\mathcal{O}(\gamma)$ at the initial point. Simply dividing~\eqref{deltaxeqH} by $H$ does not work, because $F/H\sim1/\phi$ for $|\phi|\ll1$. So it might seem like for $\gamma=0$ we have a problem in determining $\delta x''(0)$, which we need to jump to the next time step. However, \eqref{deltaxeqH} is in fact well posed even for $\gamma=0$, as can be seen by expanding $H$ and $\delta x$ in power series in $\phi$. Since $H$ only has odd powers,
\be
H(\phi)=\sum_{n=0}^\infty H_{2n+1}\phi^{2n+1} \;,
\ee
$\delta x_s$ only has even powers,
\be
\delta x_s(\phi)=\sum_{n=0}^\infty a_{2n}\phi^{2n} \;.
\ee
Plugging in these two expansions into~\eqref{deltaxeqH} gives one algebraic equation from each order in $\phi$, which determines the coefficients $a_n$ in terms of $H_n$. We find in particular 
\be
\delta x_s''(\phi=0)=-\frac{1}{4}F''(0) \;.
\ee
Using Mathematica, it is straightforward to calculate many coefficients. It might therefore be tempting to solve~\eqref{deltaxeqH} entirely using these expansions, without any numerical integration. However, we need $\delta x'$ at $\phi\to\infty$, so we would need to resum this series, regardless of how many coefficients we manage to calculate.  
Although there are methods to resum series based on a finite number of coefficients, we will not do so here. We will instead use the first couple of expansion coefficients to take the first time step, from $\phi=0$ to $\phi=\Delta\phi$. For a low-order integration step we only need $\delta x_s(0)=1$, $\delta x_s'(0)=0$ and $\delta x_s''(0)$,
\be\label{deltaxphi2}
\delta x_s(\Delta\phi)\approx1-\frac{F''(0)}{8}\Delta\phi^2 \;.
\ee
We thus take the first time step analytically, and then we solve~\eqref{deltaxeqH} numerically as usual, along the real axis starting at $\phi=\Delta\phi$ with initial conditions given by~\eqref{deltaxphi2}. By adding higher powers of $\phi$ to~\eqref{deltaxphi2} we would be able to choose a larger $\Delta\phi$. However, since we only need~\eqref{deltaxphi2} for a single time step, it is simpler to just choose a sufficiently small $\Delta\phi$ so that we can use~\eqref{deltaxphi2} without adding higher-order terms. In fact, for sufficiently small $\Delta\phi$ we could simply choose $\delta x_s(\Delta\phi)\approx1$. The time step and integration order we use for the subsequent numerical integration are independent of the first, analytical step. Thus, $\delta x_s$ is to leading order independent of $\gamma$.

From~\eqref{dxaFromdxs} we find
\be
\frac{\delta x_a'(\infty)}{\delta x_s'(\infty)}\approx\int_{i\gamma/2}^\infty\frac{\ud\phi}{H\delta x_s^2} \;,
\ee
where we have put $\gamma\to0$ everywhere except in the lower integration limit, since there it is needed because of the singular integrand. To find an approximation we will subtract a simple integrand, $I(\phi)$, with the same singularity. Since $H\approx\phi$ and $\delta x_s\approx1$, we should have $I(\phi)\approx1/\phi$ for $\phi\to0$. But we cannot simply choose $I(\phi)=1/\phi$ because then $I(\phi)$ would not decay fast enough at $\phi\to\infty$. Instead we will choose $I=1/(\phi[1+a\phi])$ where $a$ is an arbitrary constant. We have
\be
\begin{split}
\int_{i\gamma/2}^\infty\frac{\ud\phi}{\phi(1+a\phi)}&=-\ln\left(a\frac{i\gamma}{2}\right)+\ln\left(1+a\frac{i\gamma}{2}\right) \\
&=-\ln\left(a\frac{\gamma}{2}\right)-\frac{i\pi}{2}+\mathcal{O}(\gamma) \;.
\end{split}
\ee
Thus,
\be\label{ddeltaxRatio}
\begin{split}
\frac{\delta x_a'(\infty)}{\delta x_s'(\infty)}\approx&-\ln\left(a\frac{\gamma}{2}\right)-\frac{i\pi}{2}\\
&+\int_0^\infty\ud\phi\left(\frac{1}{H\delta x_s^2}-\frac{1}{\phi(1+a\phi)}\right) \;.
\end{split}
\ee
This result is independent of $a$. The integral is real for real $a$, so $\text{Im}[\delta x_a'(\infty)/\delta x_s'(\infty)]\approx-i\pi/2$. If one chooses $a=\lim_{\phi\to\infty}H[(\ud/\ud\phi)\delta x_s]^2$ then the integral converges faster at $\phi\to\infty$.
Thus, since $\delta x_s$ is independent of $\gamma$ to leading order, $\delta x_a'(\infty)$ increases as $\ln(1/\gamma)$. And from~\eqref{dLocalNonlocal} and~\eqref{deltaxWronskiansLCF} we finally find
\be
d_{P,\perp}\approx \left|c_1\ln\left[\frac{1}{\gamma}\right]+c_2 \right|
\qquad
d_{\Delta,\perp}\approx\frac{c_3}{\gamma}\;,
\ee
where the constants $c_i$ are obtained by solving~\eqref{deltaxeqH} and performing the integral in~\eqref{ddeltaxRatio}.

\section{Slow convergence as $u\to\infty$ for $\gamma\ll1$}

As mentioned in the main text, for $\gamma\ll1$, we need to integrate up to very large $r$ to see convergence. We will explain why this can be expected here. One might expect that the convergence would be faster for a field which decays faster asymptotically. For example, one might expect a Gaussian pulse to lead to a relatively fast convergence. However, even for a Gaussian pulse, the convergence is not as fast as one might have expected.

As mentioned below~\eqref{g1minusg2}, we can without loss of generality choose $g(t)$ such that it has no terms that go like $a+bt+ct^2$ for $t\to\infty$. We would find the same result anyway, but this choice makes the notation somewhat simpler. With this choice, we have for a Gaussian pulse, $G'''(x)=e^{-x^2}$,
\be
G(x)=\frac{x}{4}e^{-x^2}-\frac{\sqrt{\pi}}{8}(1+2x^2)\text{erfc}(x) \;.
\ee
Both terms decay as $e^{-x^2}$ asymptotically, which seems promising for the numerical convergence. However, for $\gamma\ll1$, the instanton follows an almost light-like trajectory in the acceleration region, where $\theta$ is very small, see~\eqref{thetaA}. So, while $\theta$ eventually grows linearly in $u$ as in~\eqref{thetaAsymptotic}, it takes a very long time before $\theta$ becomes so large that $G(\theta)$ can be approximated by its asymptotic limit. In the semi-asymptotic region, where $\phi$ is large but $\theta$ is not, we can drop the exponentially suppressed terms, $G(2\phi)$ and $G'(2\phi)$, in~\eqref{Fdef}, so
\be
F\approx\frac{6}{(2\phi-\theta)^3}[(2\phi-\theta)G'(\theta)+2G(\theta)] \;.
\ee
In this region, $F=\mathcal{O}(1/\phi^2)$ is only quadratically rather than exponentially small, even if we have chosen an exponentially decaying $G$. 

\section{Perturbative limit}

In the previous sections we have derived approximations for $\gamma\ll1$. It is probably possible to derive approximations of the saddle-point approximation for $\gamma\gg1$ too, but we expect that the saddle-point approximation breaks down in this limit, so the result would then be an approximation of an approximation that is no longer valid. However, not being able to use the saddle-point method for $\gamma\gg1$ would not be a problem, because for $\gamma\gg1$ we anyway expect the probability to become perturbative, which might not be what one wants to have if one is mainly interested in the Schwinger mechanism.

However, while the saddle-point approximation of the prefactor might break down, previous studies of other processes~\cite{Torgrimsson:2017pzs,Dinu:2017uoj,Torgrimsson:2019sjn} suggest that the approximation of the exponent can still be valid, which means we can make a completely independent check of the saddle-point result for the exponent by comparing with the perturbative result. We will show that this is also the case here for fields with poles, such as the Lorentzian pulse.

When treating the field in perturbation theory, it is natural to use the Fourier transform. For the e-dipole we have
\be
{\bf Z}(x)=\int\frac{\ud^4 k}{(2\pi)^4}e^{-ikx}{\bf Z}(k) \;,
\ee
where
\be
{\bf Z}(k)=-\frac{3\pi^2E}{|{\bf k}|k_0^3}[\delta(|{\bf k}|-k_0)-\delta(|{\bf k}|+k_0)]f(k_0){\bf e}_3
\ee
and
\be
f(k_0)=\int\ud t\, e^{ik_0t}g'''(t) \;.
\ee
For the Gaussian pulse, $g'''(t)=e^{-(\omega t)^2}$ we have
\be\label{fourierGauss}
f(k_0)=\frac{\sqrt{\pi}}{\omega}\exp\left\{-\frac{k_0^2}{4\omega^2}\right\} \;,
\ee
and for the Lorentzian pulse, $g'''(t)=1/(1+[\omega t]^2)$, we have
\be\label{fourierLorentz}
f(k_0)=\frac{\pi}{\omega}\exp\left\{-\frac{|k_0|}{\omega}\right\} \;.
\ee
The exponential suppression of the probability comes from the exponential suppression of the Fourier transform at frequencies much higher than $\omega\ll1$.
Since the Fourier photons are on shell, we need to absorb at least two photons. The dominant contribution to the integrated probability comes from pairs produced at rest, ${\bf p}={\bf p}'=0$. From energy-momentum conservation, we therefore consider the absorption of $n$ photons with $4$-momentum $\{k_0,{\bf k}\}$ and $n$ photons with $\{k_0,-{\bf k}\}$, where $k_0=|{\bf k}|=1/n$ so that the sum of all the photon energies is equal to the energy of the pair, i.e. $2$ (recall $m=1$). For the Lorentzian pulse we then have
\be
\mathbb{P}_n\sim|f^{2n}(k_0)|^2\sim E^{4n}\exp\left\{-\frac{4nk_0}{\omega}\right\}=E^{4n}\exp\left\{-\frac{4}{\omega}\right\} \;.
\ee
Since the exponent is the same for all $n$, the scaling of the prefactor with $E^{4n}$ implies that the dominant contribution comes from the absorption of only two photons,
\be\label{perturbativeLorentz}
\mathbb{P}\sim E^4\exp\left\{-\frac{4}{\omega}\right\} \;.
\ee
The reason is that, while an exponential suppression as in~\eqref{fourierLorentz} might naively seem like a fast decay, it is actually a wide distribution in this context. 
Note that this exponential scaling comes from the poles of the field. It is therefore a general result for fields with poles. For example, for a Sauter pulse, $g'''(t)=\text{sech}^2(\omega t)$, we have
\be
f(k_0)=\frac{\pi k_0}{\omega^2}\sinh^{-1}\left(\frac{\pi k_0}{2\omega}\right)\approx
\frac{2\pi k_0}{\omega^2}\exp\left(-\frac{\pi k_0}{2\omega}\right) \;.
\ee

Contrast this with the Gaussian pulse~\eqref{fourierGauss}, for which we have
\be
\mathbb{P}_n\sim|f^{2n}(k_0)|^2\sim E^{4n}\exp\left\{-\frac{1}{n\omega^2}\right\} \;.
\ee
Here the exponential suppression decreases as the number of absorbed photons increases. As shown in~\cite{Torgrimsson:2017pzs}, since the prefactor still favors absorption of fewer photons, the dominant contribution to the probability comes from some dominant order $n_{dom}$ and from $n$ close to $n_{dom}$. Since $n_{dom}$ can be quite large, this means, while the probability is ``simply'' perturbative, actually calculating it might be quite challenging since one would need to consider the absorption of many photons.  

For fields with poles, such as the Sauter and Lorentzian pulses, we can also obtain $\gamma\gg1$ approximations of the widths. The perurbative amplitude to produce a pair by absorbing two Fourier photons from the field is proportional to
\be
M=\int\ud^3k\ud^3k'f(k_0)f(k'_0)(2\pi)^4\delta^4(k+k'-p-p')\dots 
\ee
If the pole closest to the real axis is $t=i\nu$, then
the Fourier transform is proportional to $f(k_0)\propto e^{-\nu k_0}$ and
\be
f(k_0)f(k'_0)\propto e^{-\nu(k_0+k'_0)}=e^{-\nu(p_0+p'_0)} \;.
\ee
For ${\bf p}^2\ll1$ and ${\bf p}^{\prime2}\ll1$ we find
\be
|M|^2\propto e^{-4\nu-\nu({\bf p}^2+{\bf p}^{\prime2})}=
e^{-4\nu-2\nu{\bf P}^2-\frac{\nu}{2}\Delta{\bf p}^2} \;.
\ee
Thus, the widths become isotropic in this limit, where
\be\label{dlargeGammaApprox}
d_P=\frac{1}{\sqrt{2\nu}}
\qquad
d_\Delta=\sqrt{\frac{2}{\nu}} \;.
\ee
For a Lorentzian pulse we have $\nu=1/\omega$ and hence $d_P=\sqrt{E\gamma/2}$ and $d_\Delta=\sqrt{2E\gamma}$. Agreement with the numerical results is demonstrated in Fig.~\ref{fig:widthsLorentz}. \eqref{dlargeGammaApprox} has been derived for fields with poles, and so does not apply to the Gaussian field. We can see in~\eqref{fig:comparison} that we nevertheless have $d_{P,\perp}\approx d_{P,z}$ and $d_{\Delta,\perp}\approx d_{\Delta,z}$ also for the Gaussian field, but the convergence of the ratio $d_\Delta/d_P$ seems very slow.

\section{Time-dependent-field approximation}

An e-dipole field is an exactly solution to Maxwell's equations. Given a choice of pulse function, $g$, we only have two parameters to tune, $E$ and $\gamma$ (or $\omega$). We can make the field faster or slower by tuning $\gamma$, but we cannot independently make e.g. the $z$ dependence slower without also making the $t$ dependence slower. One might therefore wonder whether a purely time dependent electric field can ever be used as an approximation for these fields. But we saw in the previous section that for $\gamma\gg1$ we can use perturbation theory where the dominant contribution comes from absorbing photons such that the sum of the spatial components of the photon momenta vanish. The exponential part of the probability is then the same as what one would have if the absorbed photons were off shell with ${\bf k}=0$ rather than on shell. Such off-shell photons would be possible for a purely time-dependent field $E(t)$. For $E(t)$ one can produce a pair by absorbing a single photon. For example, for a Lorentzian pulse, $E(t)=E_0/(1+[\omega t]^2)$, we have (cf.~\cite{Popov:1972})
\be\label{perturbativeLorentzTime}
\mathbb{P}\sim E^2\exp\left\{-\frac{4}{\omega}\right\} \;.
\ee
While the prefactor is different, the exponent is exactly the same as~\eqref{perturbativeLorentz}. For a Gaussian pulse it would be much harder to calculate the perturbative result since one would need to consider the absorption of many photons. But the possibility that the result would be similar to a result for a Gaussian $E(t)$, suggests that we compare our instanton results for the e-dipole field with the corresponding instanton (or WKB) result for $E(t)$.

For $E(t)$ there is a compact result for a general pulse shape (assuming symmetry and a single maximum), see~\cite{Popov:2005,Dunne:2006st}. We write the field as $E(t)=A'(t)$ and $A(t)=f(\omega t)/\gamma$. 
The exponential part of the probability is given by
\be
\mathbb{P}\approx...\exp\left\{-\frac{\pi}{E}\bar{g}(\gamma)\right\} \;,
\ee
where $\bar{g}(\gamma)$ (which should not be confused with the dipole function $g$) is given by
\be\label{barg}
\bar{g}(\gamma)=\frac{4}{\pi\gamma^2}\int_0^{v_1}\ud v\sqrt{\gamma^2-\tilde{f}^2(v)} \;,
\ee
where $\tilde{f}(v)=-i f(iv)$, and $v_1$ is the point where $\tilde{f}(v)=\gamma$. The integral is real since $f$ is an antisymmetric function. For example, for the Lorentzian pulse we have $f(v)=\text{arctan}(v)$ and $\tilde{f}(v)=\text{arctanh}(v)$.

If $\tilde{f}(v)$ has a pole at $v_p$, then for $\gamma\gg1$
\be
\exp\left\{-\frac{\pi}{E}\bar{g}(\gamma)\right\}\approx
\exp\left\{-\frac{4v_p}{\omega}\right\} \;,
\ee
which agrees with the perturbative result, e.g.~\eqref{perturbativeLorentz} for the Lorentzian pulse.

For $\gamma\ll1$ we can Taylor expand, and we find for an arbitrary pulse shape
\be\label{gLCF}
\bar{g}(\gamma)=1-\frac{\gamma^2}{4}+\frac{40-f^{(5)}(0)}{192}\gamma^4+\mathcal{O}(\gamma^6) \;,
\ee
where we have normalized the field so that
\be\label{fTimeNorm}
f'(0)=1 \qquad f'''(0)=-2 \;.
\ee
Compare this with the corresponding result for e-dipole fields~\eqref{AedipoleNLO}. To compare we choose $E(t)=EG'''(\omega t)$, so $f(u)=G''(u)$ and in particular $f^{(5)}(0)=G^{(7)}(0)$.

\begin{figure}[!ht]
\centering
\includegraphics[width=\linewidth]{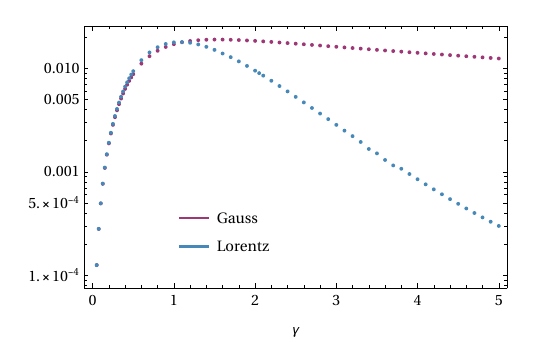}
\caption{Relative error~\eqref{errortdep} between the exponents of the exact result for the 4D dipole pulse and the purely time-dependent field $E(t) =E g'''(t)$ for the Gaussian and Lorentzian shape.}
\label{fig:timedependent}
\end{figure}

In Fig.~\ref{fig:timedependent} we see that $\mathcal{A}$ for the e-dipole field does indeed seem to converge to $\mathcal{A}$ for $E(t)$ as $\gamma$ increases. In fact, we see that the result for $E(t)$ is actually a decent approximation for all values of $\gamma$. Since all results agree on $\mathcal{A}(\gamma=0)=\pi/E$, one can expect a maximum relative error, 
\be\label{errortdep}
\epsilon=\left|\frac{\mathcal{A}[E(t)]}{\mathcal{A}[\text{e-dipole}]}-1\right| \;,
\ee
somewhere around $\gamma\sim1$. This is indeed what we find, but the maximum $\epsilon$ is only $\lesssim0.02$. This is interesting because when one sees such a small difference, the first guess would be that it is due to the smallness of some parameter. But that is not the case here, because $\mathcal{A}$ only depends on $\gamma$, and $\gamma\sim1$ is neither small nor large. The reason for the small $\epsilon$ is instead due to the fact that the functional form of $\mathcal{A}[E(t)]$ and $\mathcal{A}[\text{e-dipole}]$ are similar. They both start at $1$ for $\gamma=0$ and converge for $\gamma\gg1$, and, since they are both monotonically decreasing, there is not much that could happen in the region between $\gamma\ll1$ and $\gamma\gg1$.  
Compare the expansions in $\gamma\ll1$ for $E(t)$ in~\eqref{gLCF} and for an e-dipole in~\eqref{AedipoleNLO}. They are both power series in $\gamma^2$ and the NLO has the same sign. The coefficients, $1/4$ and $1/5$, are different but happens to be quite close. If we tried to improve the agreement by rescaling $\gamma\to\sqrt{4/5}\gamma$ for $\mathcal{A}[E(t)]$ then $\epsilon$ would become smaller for $\gamma\ll1$, but we would introduce a relatively large discrepancy at $\gamma\gg1$ on the order of $|\sqrt{4/5}-1|=\mathcal{O}(0.1)$.     

Given this agreement between $\mathcal{A}[\text{e-dipole}]$ and $\mathcal{A}[E(t)]$, it might be tempting to go beyond the leading order and treat the $z$ dependence and to consider the prefactor too. However, there are fundamental differences for the prefactor. For example, for $E(t)$ there are volume factors, which we do not have for 4D fields, and 4D fields have more nonzero and independent widths.

\section{Widths for 2D and 1D fields}\label{Widths for 2D and 1D fields}

In this section we explain to what extent results for the widths for 4D fields can, or rather cannot, be approximated by considering 2D or 1D fields. There is no parameter in the e-dipole field that we can tune such that the field becomes slower and slower in the transverse $x^\LCperp=\{x,y\}$ directions. Indeed, a field given entirely in terms of a longitudinal electric field, $E_3(t,z)$, is not a solution to Maxwell's equations (without a current). We will therefore artificially make the $x^\LCperp$ dependence slower by e.g. rescaling $x^\LCperp\to\epsilon x^\LCperp$ in the gauge potential $A_\mu$. The resulting field will no longer be a solution to Maxwell's equations, but neither are the 2D and 1D fields we want to compare with. In the 2D limit the equations for the longitudinal widths stay the same. But for the transverse widths we have (cf.~\eqref{deltaxeq})
\be\label{deltaxeq2D}
\begin{split}
    \delta x'' &= (t' \pa_x E_x - z' \pa_x B_y) \, \delta x \ne-\frac{1}{2}\nabla E\cdot\{z',t'\}\delta x \;.
\end{split}
\ee
In the 4D case we used Maxwell's equations to rewrite this equation in terms of the $\nabla E$ term, but that is not possible here. After rescaling $x^\LCperp\to\epsilon x^\LCperp$ we have
\be
\delta x''(u)=R(u)\delta x(u) \;,
\ee
where $R=\mathcal{O}(\epsilon^2)\ll1$. To leading order we have $\delta x_a\approx u$ and hence~\eqref{dFromW} gives
\be\label{dP2D}
d_{P,\perp}^{-2}\to-2\text{Im }u_c \;,
\ee
which agrees with Eq.~(104) in~\cite{DegliEsposti:2022yqw} (which simplifies using our preferred $u$ contour). The symmetric solution is more nontrivial,
\be
\delta x_s(u)\approx1+\delta x_s^{(1)}(u) 
\qquad
\delta x_s^{(1)\prime}(u)=\int_0^u\ud u'\, R(u') \;,
\ee
and~\eqref{dFromW} gives
\be
d_{\Delta,\perp}^{-2}\to\frac{1}{2}\frac{\text{Im }\delta x_s^{(1)\prime}}{|\delta x_s^{(1)\prime}|^2}=-\frac{1}{2}\text{Im}\frac{1}{\delta x_s^{(1)\prime}} \;.
\ee
Thus, $d_{\Delta,\perp}\to0$ in the limit $\epsilon\to0$. This is expected since if we had instead started with a field that does not depend on $x^\LCperp$, then we would have had momentum conservation $\delta^2(p_\LCperp+p'_\LCperp)$, and $d_{\Delta,\perp}$ gives the width for $\Delta p_\LCperp=p_\LCperp+p'_\LCperp$. For a nonzero $\epsilon\ll1$ we therefore have a regularized delta function. For the prefactor we also need
\be
|\bar{\phi}'|=2|\delta x_s'\delta x_a'|\to 2|\delta x_s^{(1)\prime}| \;,
\ee
so for the integrated probability we have (considering only those factors that involve $\epsilon$)
\be
\frac{1}{|\bar{\phi}'|^2d_{\Delta,\perp}^{-2}}\to\frac{1}{2\text{Im }\delta x_s^{(1)\prime}} \;.
\ee
The prefactor hence scales as $1/\epsilon^2$. This is also expected, because had we started with a 2D field we would have had a transverse volume factor, $V_\LCperp=V_x V_y$, so $1/\epsilon^2$ provides a regularized volume factor. 

Thus, if one starts with a 2D field, one has a constant volume factor $V_\LCperp$ and $d_{\Delta,\perp}=0$. One cannot use these trivial results to approximate anything. Judging from the 2D results, one might have wondered if perhaps $d_{\Delta,\perp}$ is at least in some sense small in the 4D case. However, Figs.~\ref{fig:comparison} and~\ref{fig:widthsLorentz} show that $d_{\Delta,\perp}$ is on the same order of magnitude as the longitudinal widths.   

Next we go one step further and take the limit where also the $z$ dependence becomes very slow. We showed in~\cite{DegliEsposti:2022yqw} that $d_{\Delta,z}\to0$, consistent with the fact that for a purely time dependent field we would have momentum conservation in all spatial directions, $\delta^3({\bf p}+{\bf p}')$. We also checked in~\cite{DegliEsposti:2022yqw} that, in the case of a Sauter pulse $E_3(t)=E/\cosh^2(\omega t)$, the two nonzero widths agree with the results in~\cite{Popov:1972}. Now we will check this for an arbitrary pulse shape (but still assuming a symmetric field with a single maximum).
For $E_3(t)=A'(t)$, $A(t)=f(\omega t)/\gamma$, we have 
\be
t'(u)=\sqrt{1+A^2(t)} \;,
\ee
which we can use to change integration variable from $u$ to $t$. For example,
\be
u_c=\int_0^{u_c}\ud u=\int_{t(0)}^0\frac{\ud t}{t'}=-\int_0^{\tilde{t}}\frac{\ud t}{\sqrt{1+A^2(t)}} \;,
\ee
where $A(\tilde{t})=i$, so from~\eqref{dP2D} we find
\be
d_{P,\perp}^{-2}\to 2\text{Im}\int_0^{\tilde{t}}\frac{\ud t}{\sqrt{1+A^2(t)}}=:\frac{\pi}{E}B_2(\gamma) \;.
\ee
Since the integration goes along the imaginary axis, we change variable and rewrite the field in terms of $\tilde{f}(v)=-i f(iv)$,
\be\label{B2v}
B_2=\frac{2}{\pi\gamma}\int_0^{v_1}\frac{\ud v}{\sqrt{1-(\tilde{f}(v)/\gamma)^2}} \;,
\ee
where $\tilde{f}(v_1)=\gamma$. This integral is similar to~\eqref{barg}. To compare with the results in~\cite{Popov:2005} we change variable from $v$ to $x=\tilde{f}(v)/\gamma$. For the Jacobian we have $\tilde{f}'(v)=F(\tilde{f}(v))$, where $F$ is some function that depends on the choice of field. For example, for a Sauter pulse we have $\tilde{f}(v)=\tan(v)$ and $F=1+\tilde{f}^2$. We find
\be\label{B2x}
B_2(\gamma)=\frac{2}{\pi}\int_0^1\frac{\ud x}{F(\gamma x)\sqrt{1-x^2}} \;.
\ee
With the same change of variable, $\bar{g}$ in~\eqref{barg} becomes
\be\label{bargx}
\bar{g}=\frac{4}{\pi}\int_0^1\frac{\ud x}{F(\gamma x)}\sqrt{1-x^2} \;.
\ee
\eqref{bargx} and~\eqref{B2x} agree with Eq.~(7.5) in~\cite{Popov:2005}.

For $d_{P,z}$ we need to solve (cf.~\eqref{deltaxeq})
\be\label{eta1D}
\eta''=[E^2(t)+E'(t)z']\eta \;.
\ee
Since one solution to~\eqref{eta1D} is $\eta=t'$, we can use Abel's identity and write the solution with correct initial conditions as
\be
\eta_s(u)=t'(u)\left(a+b\int_{u_c}^u\frac{\ud v}{t^{\prime2}(v)}\right) \;,
\ee
where $a$ and $b$ are two constants. Since the initial conditions~\eqref{initialConditionsetadelta} are set at $u=0$, and $t'(0)=0$, we have a singular integrand. However, we only need $\eta_s(u)$ for $r>0$, so we never have to integrate over $u=0$, and the limit $u\to0$ is finite,
\be
\eta_s(0)=-\frac{b}{t''(0)}\overset{!}{=}1 \;,
\ee
so $b=-t''(0)$. The Lorentz-force equation and partial integration gives
\be
\int_{u_c}^u\frac{\ud v}{t^{\prime2}}=\int_{u_c}^u\frac{\ud v}{E}\frac{\ud}{\ud v}\frac{z'}{t'}=\frac{z'}{Et'}+\int_{u_c}^u\ud v\frac{z'E'(t)}{E^2} \;,
\ee
which we use to simplify
\be
\eta_s'(u)=at''+bt'+bt''\int_{u_c}^u\ud v\frac{z'E'(t)}{E^2} \;.
\ee
The second initial condition, $\eta_s'(0)=0$, now implies
\be\label{aandb}
a=-b\int_{u_c}^0\ud v\frac{z'E'(t)}{E^2} \;.
\ee
For the nonlocal part of~\eqref{dFromW} we have
\be
p_0\eta_s'(\infty)=b \;,
\ee
and for the local, Wronskian part we need
\be
\eta_s(u_c)=a
\qquad
\eta_s'(u_c)=b \;,
\ee
where we have used $z'(u_c)=t''(u_c)=0$ and $t'(u_c)=1$, so
\be
d_{P,z}^{-2}=2\frac{W(\eta_{sr},\eta_{si})}{p_0^2|\eta_s'|^2}=2\frac{a_rb_i-b_ra_i}{|b|^2}=-2\text{Im}\frac{a}{b} \;.
\ee
Using~\eqref{aandb} and changing integration variable to $t$ gives
\be
d_{P,z}^{-2}=2\text{Im}\int_0^{\tilde{t}}\ud t\frac{A}{\sqrt{1+A^2}}\frac{A''}{A^{\prime2}}
=:\frac{\pi}{E}B_1(\gamma)\;.
\ee
Rewriting as in~\eqref{B2v} gives
\be
B_1=\frac{2}{\pi\gamma}\int_0^{v_1}\ud v\frac{\tilde{f}}{\sqrt{1-(\tilde{f}/\gamma)^2}}\frac{\tilde{f}''}{\tilde{f}^{\prime2}} \;.
\ee
From the definition of $F$, $\tilde{f}'(v)=F(\tilde{f}(v))$, we have $\tilde{f}''=FF'$ and
\be
\frac{\tilde{f}\tilde{f}''}{\tilde{f}^{\prime3}}=-\gamma\frac{\ud}{\ud\gamma}\frac{1}{F(\gamma x)} \;.
\ee
Thus,
\be
B_1(\gamma)=-\gamma B_2'(\gamma) \;,
\ee
which agrees with Eq.~(7.5) in~\cite{Popov:2005}.

By expanding in $\gamma\ll1$ as in~\eqref{gLCF} we find
\be\label{B12expansion}
\begin{split}
B_2&=1-\frac{\gamma^2}{2}+\left(\frac{5}{8}-\frac{f^{(5)}(0)}{64}\right)\gamma^4+\mathcal{O}(\gamma^6)\\
B_1&=\gamma^2+[f^{(5)}(0)-40]\frac{\gamma^4}{16}+\mathcal{O}(\gamma^6) \;.
\end{split}
\ee
For a monochromatic field we find agreement with the corresponding expansions in Eq.~(7.6) in~\cite{Popov:2005}, we just have to recall that with our normalization~\eqref{fTimeNorm}, we have $f(v)=\sin(\sqrt{2}v)/\sqrt{2}$, so our definition of $\gamma$ differ from that in~\cite{Popov:2005} by a factor of $\sqrt{2}$. By using the same normalization~\eqref{fTimeNorm} for all fields, we see that the two nonzero widths, $d_{P,\perp}$ and $d_{P,z}$, are to leading order independent of the pulse shape.

\section{RR}\label{RR}

To estimate the size of RR (see~\cite{Gonoskov:2021hwf} for a review) we consider the classical Landau-Lifshitz (LL) equation, 
\be\label{LLeq}
q''_\mu={F_\mu}^\nu q'_\nu+\beta\left({F'_\mu}^\nu q'_\nu+F_{\mu\nu}F^{\nu\rho}q'_\rho+[Fq']^2q_\mu'\right) \;,
\ee
where $\beta=\frac{2}{3}\frac{e^2}{4\pi}$. We consider zero transverse momenta, since the saddle point is at $p_\LCperp=p'_\LCperp=0$. 
After rescaling $F_{\mu\nu}\to E F_{\mu\nu}$, $q^\mu\to q^\mu/E$ and $u\to u/E$, \eqref{LLeq} remains the same except that $\beta\to E\beta$. RR might thus only be important if some other parameter is large enough to compensate for $E\beta\ll\beta\ll1$. We consider therefore $\gamma\ll1$. Changing variables to $\phi$ and $\theta$, and expanding to leading order in $\gamma\ll1$ gives $\phi''(u)=F(\phi)\phi'+E\beta F'(\phi)\phi''$. This is the same as the LL equation for a field given entirely by $E_3(t+z)$, which was solved in~\cite{Ekman:2021vwg,Ilderton:2023ifn}. The solution is
\be
\frac{\ud\phi}{\ud u}(\phi)=e^{E\beta F(\phi)}\int_0^\phi\ud\varphi e^{-E\beta F(\varphi)}F(\varphi) \;.
\ee
Since $F(\phi)=\mathcal{O}(1)$, there is nothing to compensate for $E\beta\ll1$, so RR is negligible. A similar conclusion and the identification of $E\alpha$ as the relevant parameter can also be found in~\cite{Bulanov:2010gb}.

Many strong-field-QED processes are studied in fields with components orthogonal to the momentum of the particles. A high-energy particle will then effectively see a much stronger field in a frame where the particle's energy is $\mathcal{O}(1)$ (this could be the rest frame for a massive particle). The field will also effectively appear as a plane wave. However, in our case, although the particles are accelerated to high energies for $\gamma\ll1$, they are accelerated along the direction of the electric field on a path where there are no transverse field components. A Lorentz boost parallel to the electric field does not change the field strength. With $E\ll1$ in the lab frame, we will therefore also have $E\ll1$ in the rest frame. Thus, rather than a plane wave, we have shown that the particle effectively sees a purely electric field which only depends on lightfront time, $E_3(t+z)$. This is not a solution to Maxwell's equations in vacuum, but that is not a problem since it does approximate an exact solution (the e-dipole field) along the relevant plane $x=y=0$. A similar point was made in~\cite{Linder:2015vta}, where it was shown that the closed worldline instanton for a standing wave, $\propto\cos(\omega t)\cos(k x)$, is the same as the instanton for a purely time-dependent electric field, $\propto\cos(\omega t)$. 

We have shown that $E_3(t+z)$ is relevant for the acceleration region for $\gamma\ll1$ because there the particles have reached highly relativistic velocities and travel along almost lightlike trajectories (see also~\cite{Brodin:2022dkd}). However, we do not approximate the field as $E_3(t+z)$ in the formation region. In fact, our results are very different from the probability of pair production by $E_3(t+z)$, which was derived in~\cite{Tomaras:2001vs,Ilderton:2014mla}. This is easy to see. The probability for $E_3(t+z)$ is proportional to volume factors in the $x$, $y$ and $t-z$ directions. We have no volume factors because we consider a 4D field.

\end{document}